\documentclass{aa}
\usepackage{graphicx}
 \newcounter{NOTE}
 \newcounter{QUESTION}
 \newcounter{NEWPART}
  \newcommand{\SKIPTEXT}[1]{}
 \renewcommand{\SKIPTEXT}[1]{{\tt \begin{center} [ ... TEXT SKIPPED ... ] \end{center}}}

 \newcommand{\CharFuncQRAData}{\Theta\Pi^{\mathrm{QRA}}_{N_{\mathrm{s}}}}
 \newcommand{\CharFuncData}{\Pi_{N_{\mathrm{s}}}}
 \newcommand{\CharFuncQRA}{\Phi_{N_{\mathrm{s}}}}

 \newcommand{\gsim}{\mbox{$\stackrel{_>}{_\sim}$}} 
 \newcommand{\lsim}{\mbox{$\stackrel{_<}{_\sim}$}} 
 \newcommand{\etal}{{et~al.}}

 \newcommand{\Ns}{\mbox{$N_{\mathrm{s}}$}}
 \newcommand{\QUANTIZE}{\mbox{$\mathcal{Q}$}}

 \newcommand{\OffsetQ}{\mbox{$\mathcal{O}_{\wr}$}}

 \newcommand{\DeltaTR}[1]{\mbox{$\Delta T_{#1}^{\mathrm{R}}$}}
 
 \newcommand{\DeltaTE}[1]{\mbox{$\Delta T_{#1}^{\mathrm{E}}$}}
 \newcommand{\EDT}[1]{\mbox{$\Delta T_{#1}^{\mathrm{E}}$}}
 \newcommand{\DeltaTstar}{\mbox{$\Delta T_{\star}$}}

 \newcommand{\QDT}[1]{\mbox{$\Delta T_{#1}^{\mathrm{Q}}$}}
 \newcommand{\RDT}[1]{\mbox{$\Delta T_{#1}^{\mathrm{R}}$}}

 \newcommand{\EXPCT}{\mbox{$\mathrm{E}$}}
 \newcommand{\mean}{\mbox{$\mathrm{mean}$}}
 
 \newcommand{\skewness}{\mbox{$\mathrm{Skewness}$}}
 \newcommand{\kurtosis}{\mbox{$\mathrm{Kurtosis}$}}

 \newcommand{\DeltaTi}{\mbox{$\Delta T_{i}$}}

 \newcommand{\fknee}{\mbox{$f_{\mathrm{knee}}$}}
 \newcommand{\Cl}{\mbox{$C_{\ell}$}}
 \newcommand{\Clwn}[1]{\mbox{$C_{\ell#1}^{wn}$}}
 \newcommand{\ClQRA}[1]{\mbox{$C_{\ell#1}^{\mathrm{QRA}}$}}
 \newcommand{\ChlQRA}[1]{\mbox{$C_{h\ell#1}^{\mathrm{QRA}}$}}
 \newcommand{\DeltaT}{\mbox{$\Delta T$}}

 \newcommand{\qADC}{\mbox{$q_{\mathrm{ADC}}$}}
 \newcommand{\floor}{\mbox{$\mathrm{floor}$}}
 \newcommand{\ceil}{\mbox{$\mathrm{ceil}$}}
 \newcommand{\round}{\mbox{$\mathrm{round}$}}
 \newcommand{\trunc}{\mbox{$\mathrm{trunc}$}}
 \newcommand{\qZero}{\mbox{$q_{\mathrm{0}}$}}

 \newcommand{\Kwn}{\mbox{$K_{\mathrm{wn}}^2$}}
 \newcommand{\Kqra}{\mbox{$K_{\mathrm{qra}}^2$}}

 \newcommand{\Npix}{\mbox{$N_{\mathrm{pix}}$}}
 \newcommand{\Keff}{\mbox{$K_{f}$}}

 \newcommand{\KurtQRA}{\mbox{$\mathrm{Kurtosis}[{\mathrm{QRA}}]$}}
 \newcommand{\KurtOrig}{\mbox{$\mathrm{Kurtosis}[{\mathrm{Orig}}]$}}

 \newcommand{\Nh}{\mbox{$N_{\mathrm{h}}$}}

  \newcommand{\alphaN}{\mbox{$\alpha_{N_{\mathrm{s}}}$}}
  \newcommand{\sigmaN}{\mbox{$\sigma_{N_{\mathrm{s}}}$}}
  \newcommand{\uN}{\mbox{$u_{N_{\mathrm{s}}}$}}
  \newcommand{\sync}{\mbox{$\mathrm{sync}$}}
  
  \newcommand{\deltatilde}{\mbox{$\tilde{\delta}_{N_{\mathrm{s}}}$}}
  \newcommand{\omegamax}{\mbox{$\omega_{\mathrm{max}}$}}

 \newcommand{\FIGUREONE}{
 \begin{figure}[t]
 \begin{center}
 \includegraphics[scale=0.36,angle=90]{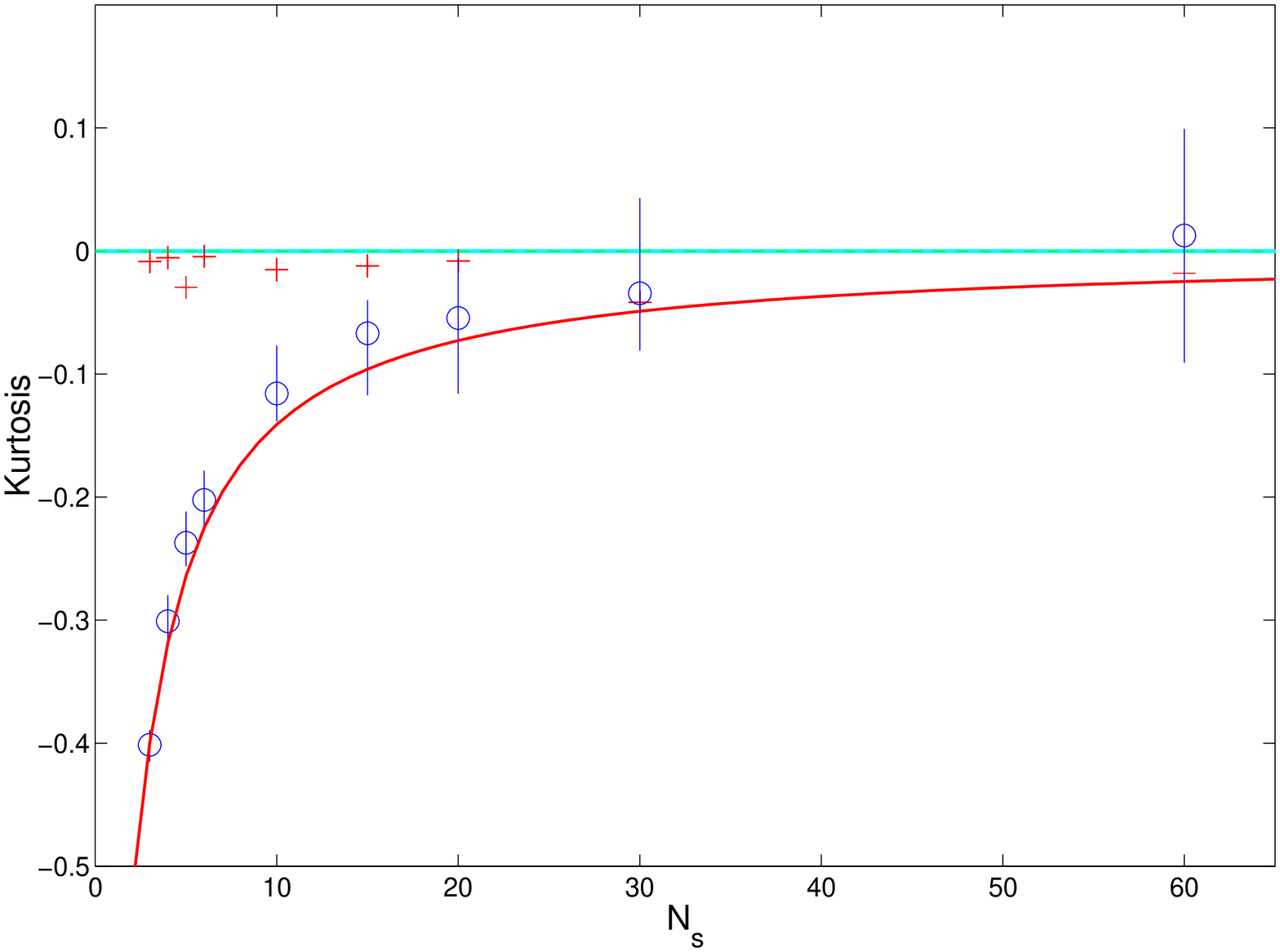}
 \end{center}
 \caption{
Kurtosis for the distribution of the random errors for the QRA
process as a function of $\Ns$. Crosses represents the kurtosis
for the original simulated samples without quantization and
reconstruction, but just averaging them in groups of \Ns. Bars
represents the distribution of values of kurtosis for the
quantization error for the averages of \Ns\ quantized and
reconstructed simulated samples. Note that the realisation of
simulated TOD is not changed when $q$ is changed so that the
variability is just due to the QRA process. Circles represent the
averaged kurtosis for the points forming the bar. The solid line
represents the theoretical expectation for the kurtosis after \Ns\
averages.
 }\label{fig:tod:kurtosis}
 \end{figure} }

 \newcommand{\FIGURETWO}{
 \begin{figure}[t]
 \begin{center}
 \includegraphics[scale=0.35,angle=90]{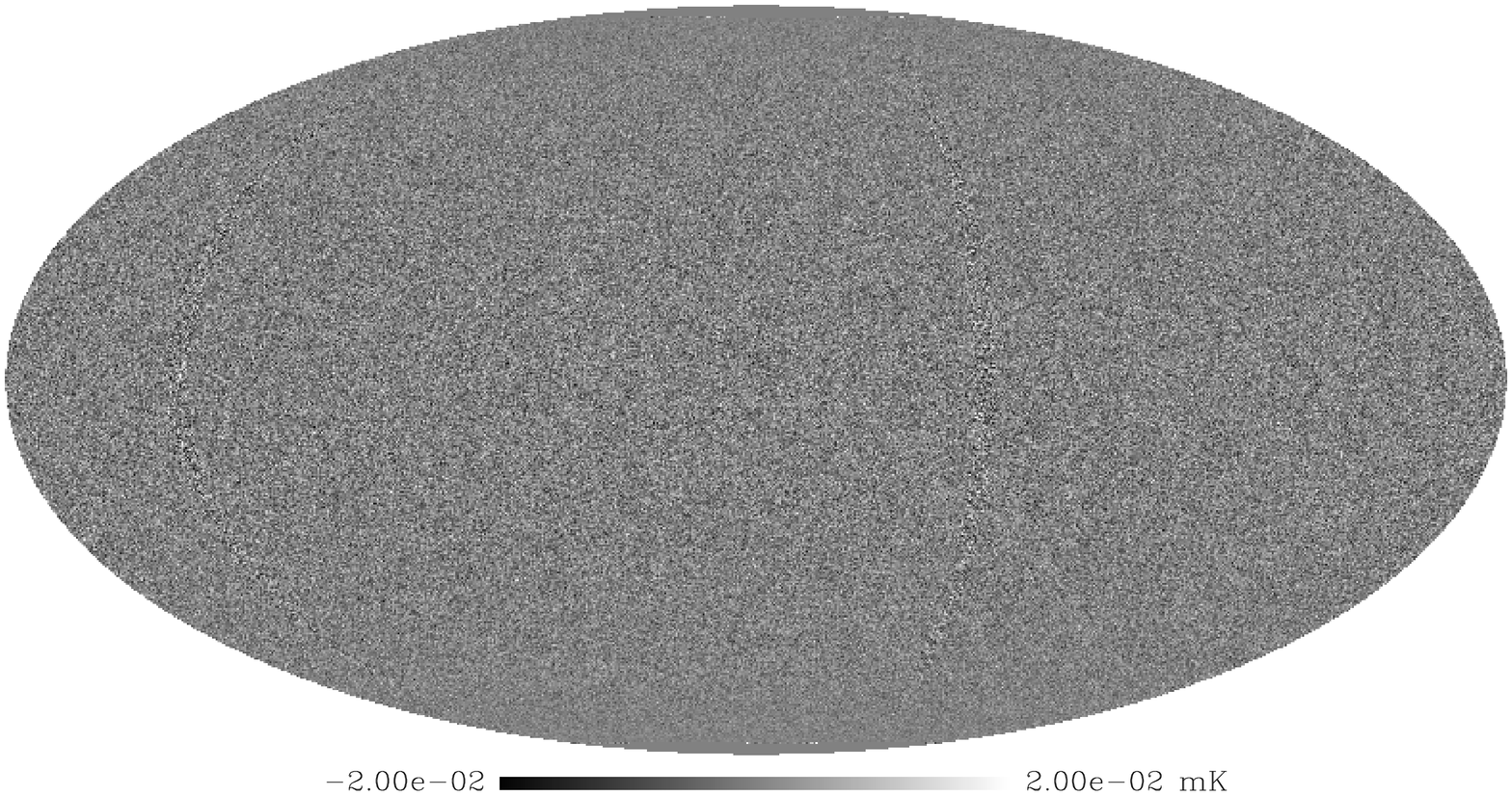}
  \vspace{0.3cm}
 \includegraphics[scale=0.35,angle=90]{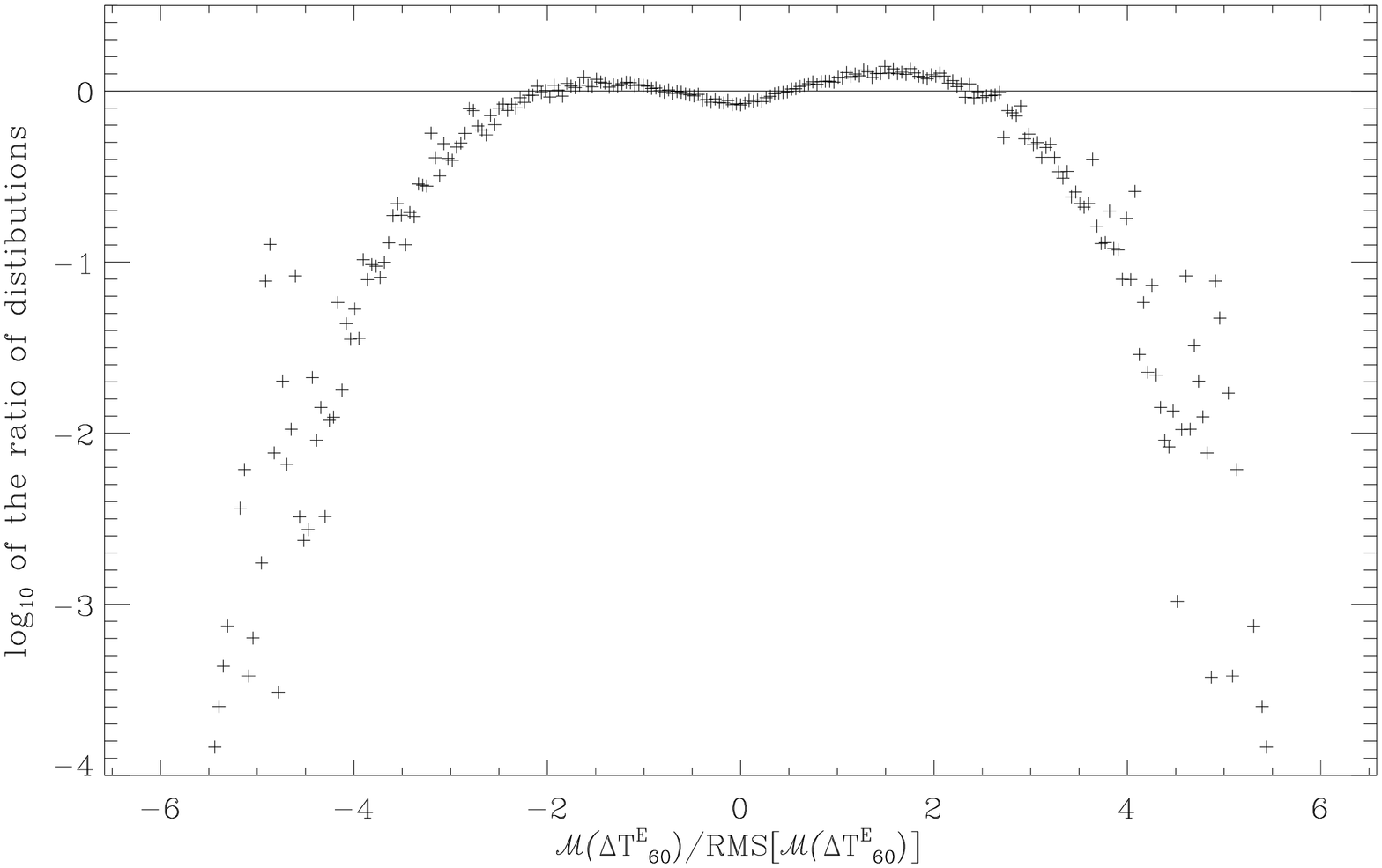}
 \end{center}
 \caption{
The upper frame is the map of the quantization error after the QRA
process, plus destriping and map making, for a single 30~GHz
radiometer, with $\sigma = 1.053$~mK, $\sigma/q \approx 2$,
$\fknee = 0.1$~Hz, $\Ns = 60$. The lower frame is the ratio of the
histogram of the map divided by the histogram of a random variable
normally distributed. The zero line represents the case of a
quantization error normally distributed. The y-scale represents
the $\log_10$\ of the ratio. As indicated even in
Tab.~\ref{tab:qerror:mapstat}\ the quantization error is not
completely normal distributed. However the largest deviations are
for large quantization errors, in excess of twice the RMS of the
QRA process.
 }\label{fig:MAP:qerror}
 \end{figure}
 }

 \newcommand{\FIGURETWOBIS}{
 \begin{figure}[t]
 \begin{center}
 \includegraphics[scale=0.35,angle=90]{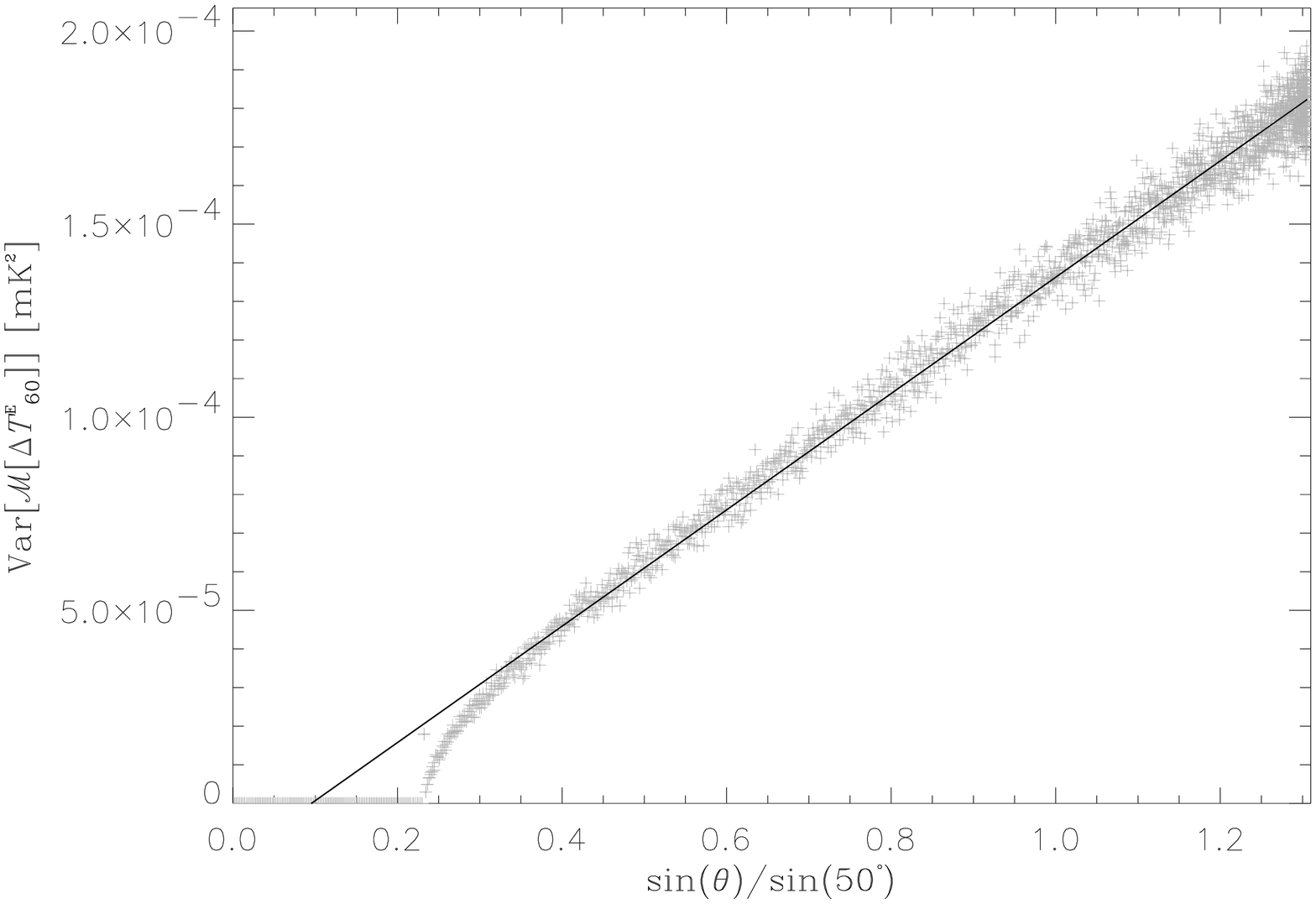}
 \end{center}
 \caption{
Colatitudinal distribution of the variance of the QRA error on a
sky map for $\sigma/q \approx 2$\ and the 100~GHz channel. The
straight line represents the best linear fit.
 }\label{fig:lat:dependence}
 \end{figure}
 }

 \newcommand{\FIGURETHREE}{
 \begin{figure}[t]
 \begin{center}
 \includegraphics[scale=0.36,angle=90]{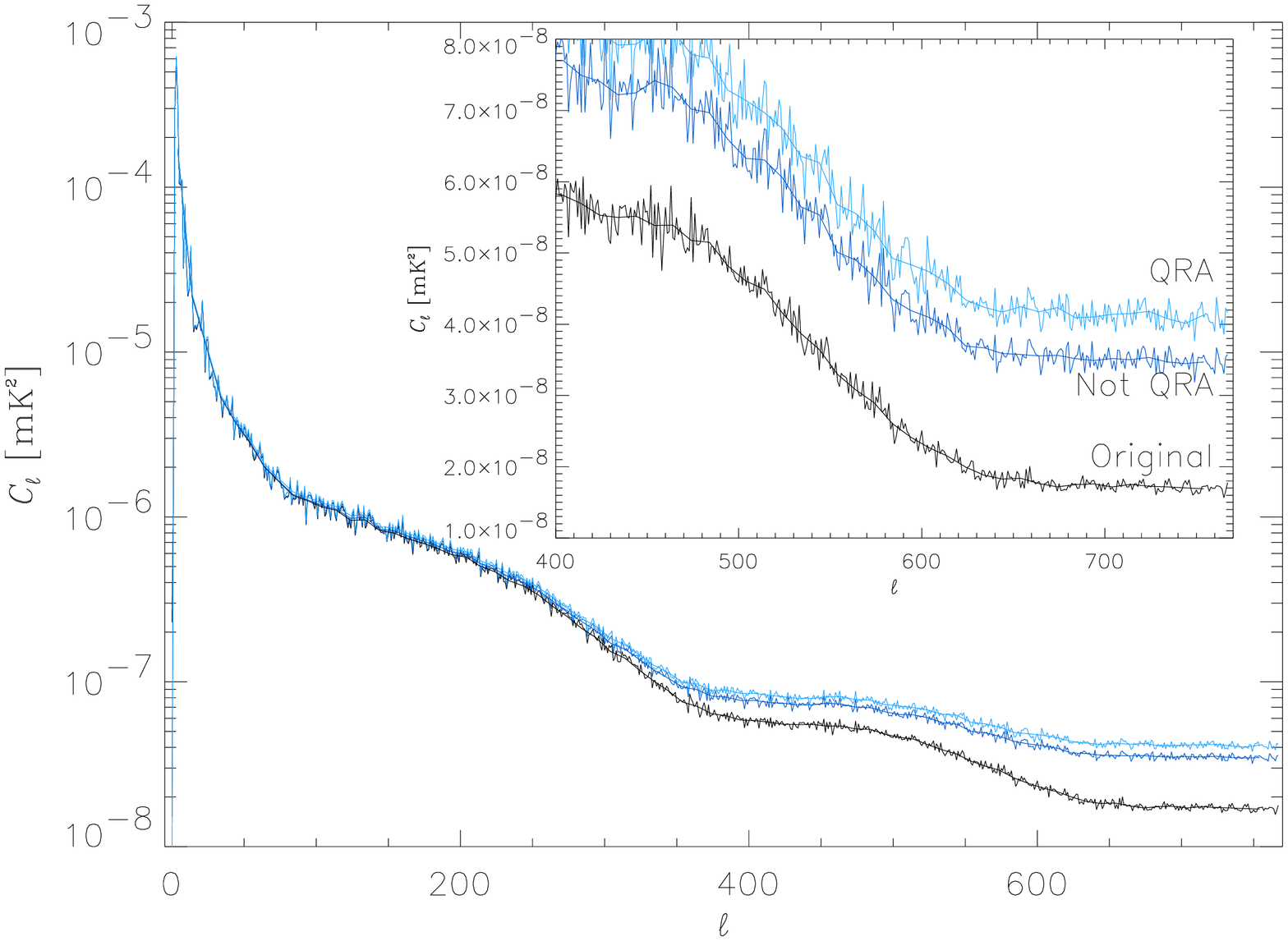}
 \end{center}
 \caption{
Power Spectra from maps for simulated CMB, simulated CMB plus
noise, simulated CMB plus noise plus QRA processing (labelled as
{\em Original}, {\em Not QRA}, {\em QRA}) for the 30~GHz channel
(1 radiometer). The inset is an enlargement of the power spectra
for $l>400$. For each power spectrum a smoothed spectrum is
overlapped to the original spectrum. The $\sigma/q \approx 0.5$
values has been chosen to increase the quantization effect. In the
nominal case $\sigma/q \approx 2$\ the effect is about 16 times
smaller and difficult to see in the plot.
 }\label{fig:pwr:spc}
 \end{figure}
 }

 \newcommand{\FIGUREFOUR}{
 \begin{figure*}[t]
 \begin{center}
 \includegraphics[scale=0.70,angle=90]{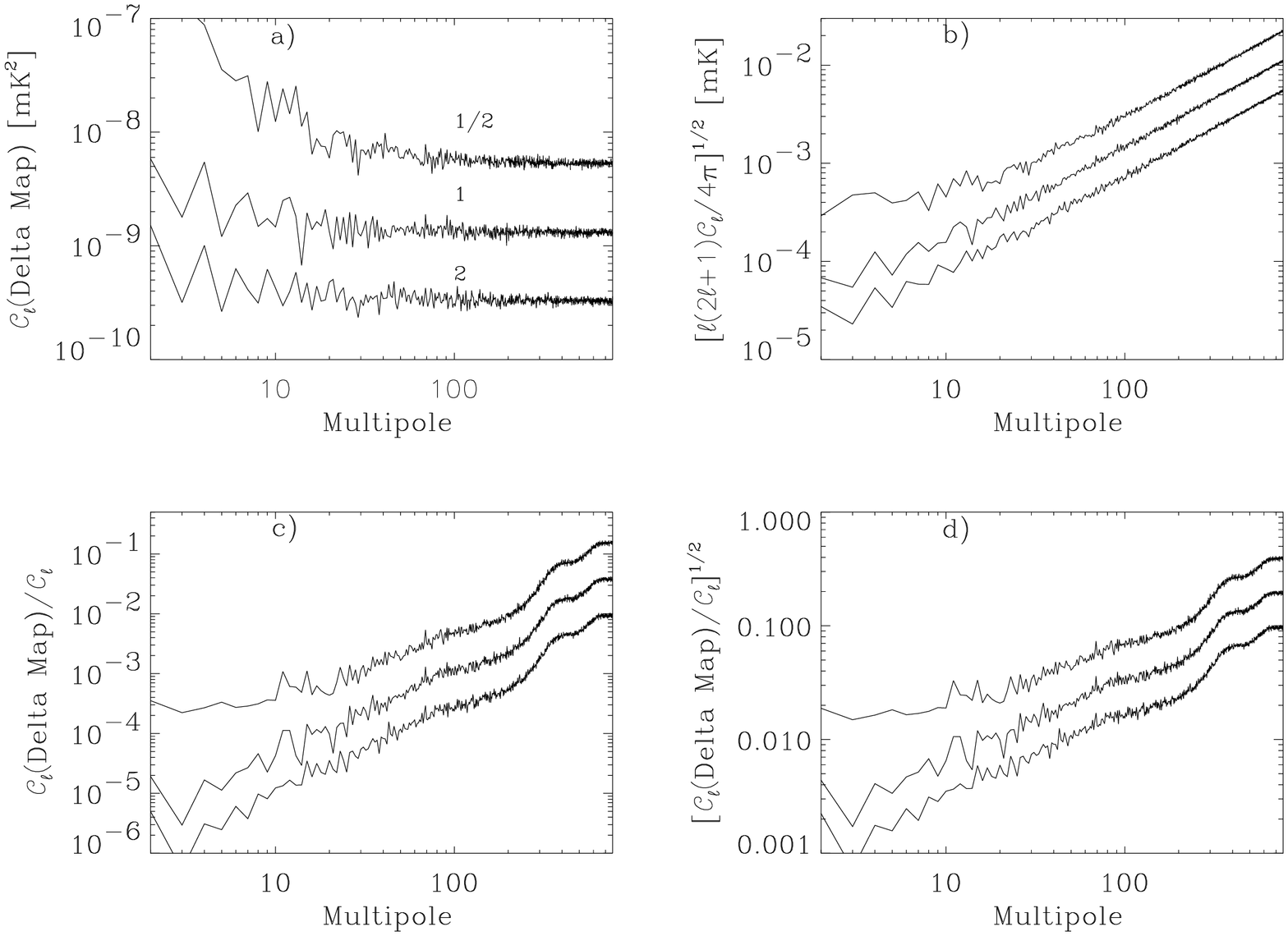}
 \end{center}
 \caption{
Power excess induced by signal quantization for the 30~GHz
channel, 1~radiometer, i.e. the power spectrum for the
quantization error map (panel a)) reported also in terms of
$\delta T_\ell = (C_\ell \ell (2\ell + 1)/4\pi)^{1/2}$ (panel b)).
Ratio between the power excess induced by the QRA process and the
power spectrum of the unquantized map (noise plus sky) (panel c))
and its square root (panel d)). The approximate value of
$\sigma/q$ for each curve is reported in panel a) (see
Tab.~\ref{tab:qerror:mapstat} for the corresponding exact values).
 }\label{fig:pwr:excess}
 \end{figure*}
 }

  \newcommand{\TABLEONE}{
 \begin{table*}[ht]
 \begin{center}
 \begin{tabular}{ccccc}
 \multicolumn{5}{c}{30~GHz} \\
 \hline
 $\sigma/q$ & \% variance increase  & \% variance increase                       & a        & b\\
            & (from simulations) &  (from eq.~\ref{eq:map:total:noise})    & (mK$^2$) & (mK$^2$)\\
 \hline
 2.076      & $\;$ 1.94  & $\;$ 1.93  &  $(-9.00 \pm 1.16) \times 10^{-7}$ & $(2.105 \pm 0.011) \times 10^{-5}$ \\
 1.038      & $\;$ 7.67  & $\;$ 7.73  &  $(-4.02 \pm 0.46) \times 10^{-6}$ & $(8.459 \pm 0.046) \times 10^{-5}$ \\
 0.519      &       30.94  &       30.59 &  $(-1.30 \pm 0.19) \times 10^{-6}$ & $(3.399 \pm 0.019) \times 10^{-4}$\\
 &&&&\\
 \multicolumn{5}{c}{100~GHz} \\
 \hline
 $\sigma/q$ & \% variance increase  & \% variance increase                       & a        & b\\
            & (from simulations) &  (from eq.~\ref{eq:map:total:noise})    & (mK$^2$) & (mK$^2$)\\
 \hline
 2.724 & $\;$ 1.13  & $\;$ 1.12 & $(-1.436 \pm 0.036) \times 10^{-5}$ & $(1.5061 \pm 0.0036) \times 10^{-4}$ \\
 1.362 & $\;$ 4.55  & $\;$ 4.49 & $(-5.903 \pm 0.146) \times 10^{-5}$ & $(6.0464 \pm 0.0145) \times 10^{-4}$ \\
 0.681 &       17.57  &       17.97 & $(-2.241 \pm 0.056) \times 10^{-4}$ & $(2.3698 \pm 0.0055) \times 10^{-3}$ \\
   \hline
 \end{tabular}
 \end{center}
 \caption{
Quantization error at map level. The table compares the \%
variance increase determined by simulations
($\left(\sigma_{\mathrm{map},\, \mathrm{total}}^2/
\sigma_{\mathrm{map},\, \mathrm{wn}}^2 - 1\right)\times 100$) with
the expectation derived from eq.~(\ref{eq:map:total:noise}). The
last two column report the best fit parameters to estimate the RMS
quantization error as a function of the pixel colatitude.
 }\label{tab:qerror:mapstat}
 \end{table*}
 }

  \newcommand{\TABLEONEB}{
 \begin{table*}[ht]
 \begin{center}
 \begin{tabular}{cccccc}
 \multicolumn{6}{c}{30~GHz} \\
 \hline
 $\sigma/q$ & $q$ & Skewness$(X)$ & Kurtosis$(X)$ & $|\Delta K / K|\%$ & $|\Delta K / K|\%$ \\
            & [mK]&               &               & (no $1/f^\alpha$)  & ($1/f^\alpha$ + dest.)\\
 \hline
 2.076      & 0.509 & -0.00029069   & 0.370293 & 30.7 & 4.2 \\
 1.038      & 1.017 &  0.00692233   & 0.394023 & 62.8 & 5.9   \\
 0.519      & 2.035 &  0.00547806   & 0.383053 & 63.9 & 26.1 \\
 &&&&&\\
 \multicolumn{6}{c}{100~GHz} \\
 \hline
 $\sigma/q$ & $q$ & Skewness$(X)$ & Kurtosis$(X)$ & $|\Delta K / K|\%$ & $|\Delta K / K|\%$ \\
            & (mK)&               &               & (no $1/f^\alpha$)           &
($1/f^\alpha$ + dest.)\\
 \hline
 2.724 & 1.221 & 0.00333290 & 0.257705 & 0.4  & 1.5 \\
 1.362 & 2.441 & 0.00031871 & 0.274681 & 3.1  & 5.4 \\
 0.681 & 4.883 & -0.00492754 & 0.267852 & 14.9 & 8.8\\
  \hline
 \end{tabular}
 \end{center}
 \caption{
Table of sampling statistics for the quantization error maps:
skewness and kurtosis for the QRA. The fifth column lists values
of $|\Delta K / K| \times 100$\ obtained comparing the kurtosis of
maps not containing $1/f^\alpha$ noise and not destriped, with and
without QRA processing. The same for the last column where
$1/f^\alpha$ noise has been introduced and destriping has been
applied. For other details see Tab.~\ref{tab:qerror:mapstat}.
 }\label{tab:qerror:mapstat:nongaus}
 \end{table*}
 }

  \newcommand{\TABLETWO}{
 \begin{table}[ht]
 \begin{center}
 \begin{tabular}{ccccccc}
 \multicolumn{7}{c}{30~GHz} \\
 $\sigma/q$ & $q$ &
 \fknee & prec. &
 \multicolumn{3}{c}{$\langle {\it C}_{\ell>70,60}^{\mathrm{exc}}\rangle$} \\
                  & (mK) & (Hz) & & \multicolumn{3}{c}{($10^{-9} mK^2$)} \\
 \hline
 2.076 & 0.509 &  0.1 & no &    0.331 & $\pm$ &        0.020 \\
 1.038 & 1.017 &  0.1 & no &    1.323 & $\pm$ &        0.081 \\
 0.519 & 2.035 &  0.1 & no &    5.410 & $\pm$ &        0.350 \\
   &&&&&\\
 \multicolumn{7}{c}{100~GHz} \\
 $\sigma/q$ & $q$ &
 \fknee & prec. &
 \multicolumn{3}{c}{$$\ClQRA{,60}$$} \\
                  & (mK) & (Hz) & & \multicolumn{3}{c}{($10^{-9} mK^2$)} \\
 \hline
 2.724 & 1.221 &    0.1 & no &    0.561 & $\pm$ &        0.027 \\
 1.362 & 2.441 &    0.1 & no &    2.242 & $\pm$ &        0.110 \\
 0.681 & 4.883 &    0.1 & no &    8.877 & $\pm$ &        0.470 \\
   &&&&&\\
 \multicolumn{7}{c}{30~GHz} \\
 $\sigma/q$ & $q$ &
 \fknee & prec. &
 \multicolumn{3}{c}{$$\ClQRA{,60}$$} \\
                  & (mK) & (Hz) & & \multicolumn{3}{c}{($10^{-9} mK^2$)} \\
 \hline
 1.038 & 1.017 &     0.5 & no  &   1.323 & $\pm$ &        0.076 \\
 0.519 & 2.035 &     0.1 & yes &   5.820 & $\pm$ &        0.350 \\
 \hline
 \end{tabular}
 \end{center}
 \caption{
Values of $\ClQRA{,60}$\ obtained from simulations for a single
30~GHz and 100~GHz horn. Errors represents $1\,\sigma$\ dispersion
of $\Cl$\ about the average. The last subtable reports the effect
of a change of \fknee\ for the baseline scanning strategy and for
a scanning strategy which includes a precession of the spin axis.
 }\label{tab:cl70}
 \end{table}
 }

  \newcommand{\TABLETHREE}{
 \begin{table*}[t]
 \begin{center}
 \begin{tabular}{rccccc}
 \multicolumn{1}{c}{
 Frequency}    & $B$ & $A$ & $A"$ &  $\ClQRA{,1}$ & $\Keff$\\
 \multicolumn{1}{c}{
 Channel}      &     & ($\log_{10} mK^2$) & ($\log_{10} mK^2$) & ($mK^2$) & \\
 \hline
  30~GHz & $+0.015  \pm 0.022$ &   $-8.8425 \pm 0.0013$  &  $-8.8903 \pm 0.0156$ & $(0.7724 \pm 0.0120) \times 10^{-7}$ & $0.7241 \pm 0.0113$\\
 100~GHz & $-0.008  \pm 0.023$ &   $-8.3834 \pm 0.0018$  &  $-9.4257 \pm 0.0125$ & $(0.2251 \pm 0.0028) \times 10^{-7}$ & $0.8364 \pm 0.0105$\\
 \end{tabular}
 \end{center}
 \caption{
Results of the fit of $\log_{10} \ClQRA{,60}$\ versus $\log_{10}
\sigma/q$\ for the data in Tab.~\ref{tab:cl70}. Only the data in
the first and the second subframes have been included in the fit.
 }\label{tab:cl70:fit}
 \end{table*}
 }

  \newcommand{\TABLEAPPONE}{
 \begin{table}[t]
 \begin{center}
 \begin{tabular}{rccc}
 $\Ns$ & $\deltatilde$ & \sigmaN & \alphaN \\
  \hline
  1 & 0.3414 & 0.2887 & 1.1825 \\
  2 & 0.2183 & 0.2041 & 1.0696 \\
  3 & 0.1726 & 0.1667 & 1.0359 \\
  4 & 0.1482 & 0.1443 & 1.0269 \\
  5 & 0.1321 & 0.1291 & 1.0230 \\
  6 & 0.1205 & 0.1179 & 1.0227 \\
 10 & 0.0930 & 0.0913 & 1.0187 \\
 15 & 0.0753 & 0.0745 & 1.0108 \\
 30 & 0.0529 & 0.0527 & 1.0033 \\
 60 & 0.0373 & 0.0373 & 1.0017 \\
 \end{tabular}
 \end{center}
 \caption{
Table of values of $\deltatilde$, \sigmaN\ and \alphaN\ as a
function of $\Ns$.
 }\label{tab:app:one}
 \end{table}
 }

 \newcommand{\SUBTABLEHEADERMOMENT}[1]{
     \multicolumn{7}{c}{#1}\\
     & \multicolumn{3}{c}{Not Convolved}
     & \multicolumn{3}{c}{Convolved} \\
   \multicolumn{1}{c}{$N_{\mathrm{side}}$} &
   \multicolumn{1}{c}{rms} &
   \multicolumn{1}{c}{skew} &
   \multicolumn{1}{c}{kurt} &
   \multicolumn{1}{c}{rms} &
   \multicolumn{1}{c}{skew} &
   \multicolumn{1}{c}{kurt} \\
   \hline
   }

 \newcommand{\TABLEAPPCONE}{
 \begin{table*}[ht]
 \begin{center}
 \begin{tabular}{rrrrrrr}
 \SUBTABLEHEADERMOMENT{CMB}
  1024 &      0.0994 &       0.0138 &        0.0683 & 0.0744 &       0.0259 &       0.0728\\
  512 &       0.0982 &       0.0149 &        0.0137 & 0.0748 &       0.0254 &       0.0202\\
  256 &       0.0930 &       0.0179 &        0.0116 & 0.0743 &       0.0259 &       0.0176\\
 \hline
 &&&&&&\\
 \SUBTABLEHEADERMOMENT{Galactic dust}
 1024 &       0.0135 &      50.5199 &    4856.6387  & 0.0112 &      27.5733 &    1439.6084\\
  512 &       0.0133 &      47.1236 &    4258.4253  & 0.0112 &      27.3251 &    1417.4308\\
  256 &       0.0128 &      41.5666 &    3382.7876  & 0.0112 &      26.7408 &    1357.6565\\
 \hline
 &&&&&&\\
 \SUBTABLEHEADERMOMENT{Galactic synchrotron}
 1024 &       0.1149 &       5.7031 &      43.5534  & 0.1147 &       5.6796 &      43.1066\\
  512 &       0.1189 &       5.3191 &      38.9364  & 0.1187 &       5.3008 &      38.5514\\
  256 &       0.1190 &       5.2726 &      38.7467  & 0.1188 &       5.2529 &      38.3544\\
 \hline
 &&&&&&\\
 \SUBTABLEHEADERMOMENT{Thermal SZ}
 1024 &       0.0289 &      -2.4063 &      11.5482  & 0.0038 &      -1.9015 &       4.3527\\
  512 &       0.0165 &      -1.2956 &       3.2362  & 0.0037 &      -1.2614 &       3.4428\\
  256 &       0.0097 &      -1.0981 &       2.6870  & 0.0036 &      -1.2332 &       3.3694\\
 \hline
 &&&&&&\\
 \SUBTABLEHEADERMOMENT{Extragalactic sources}
 1024 &       0.2877 &     142.7467 &   32287.1797  & 0.0343 &      31.0366 &    1803.6383\\
  512 &       0.2064 &     119.9874 &   23013.3711  & 0.0342 &      30.1208 &    1728.2690\\
  256 &       0.1116 &      63.1182 &    6254.4443  & 0.0334 &      28.8417 &    1591.0974\\
 \hline
 \end{tabular}
 \end{center}
 \caption{
Statistical moments (rms, skewness and kurtosis indices) of
simulated maps, component by component (maps expressed in terms of
antenna temperature in mK units).
 }\label{tab:app:c:one}
 \end{table*}
 } 

 \newcommand{\TABLEAPPCTWO}{
 \begin{table*}[ht]
 \begin{center}
 \begin{tabular}{rrrrrrr}
 \SUBTABLEHEADERMOMENT{Galactic dust, $|b| > 30^\circ$ }
 1024 &       0.0003 &     119.5806 &   40805.9648 & 0.0003 &      18.3523 &    1182.7938\\
  512 &       0.0003 &      95.2859 &   27483.1035 & 0.0003 &      18.0818 &    1152.1223\\
  256 &       0.0003 &      59.3470 &   10995.2607 & 0.0003 &      17.4658 &    1075.4803\\
 \hline
 &&&&&&\\
 \SUBTABLEHEADERMOMENT{Galactic synchrotron, $|b| > 30^\circ$}
 1024 &       0.0231 &       1.7020 &       4.6648 & 0.0228 &       1.7550 &       4.8509\\
  512 &       0.0232 &       1.6361 &       4.5589 & 0.0229 &       1.6856 &       4.7311\\
  256 &       0.0232 &       1.6407 &       4.5924 & 0.0229 &       1.6798 &       4.7297\\
 \hline
 &&&&&&\\
 \SUBTABLEHEADERMOMENT{Thermal SZ, $5 \sigma\mathrm{-clipping}$}
 1024 &       0.0289 &      -2.3803 &      11.0141 & 0.0038 &      -1.9015 &       4.3527\\
  512 &       0.0165 &      -1.2956 &       3.2362 & 0.0037 &      -1.2614 &       3.4428\\
  256 &       0.0097 &      -1.0981 &       2.6870 & 0.0036 &      -1.2332 &       3.3694\\
 \hline
 &&&&&&\\
 \SUBTABLEHEADERMOMENT{Extragalactic sources, $5 \sigma\mathrm{-clipping}$}
 1024 &       0.0351 &       7.0355 &      63.6241 & 0.0197 &       7.9949 &      89.2986\\
  512 &       0.0332 &       6.5662 &      57.6648 & 0.0199 &       7.4744 &      83.2827\\
  256 &       0.0307 &       6.8692 &      64.2773 & 0.0198 &       7.4770 &      84.4513\\
 \hline
 \end{tabular}
 \end{center}
 \caption{
Statistical moments (rms, skewness and kurtosis indices) of
simulated maps, component by component. Several ``masks'' are
considered: simple Galactic cuts in the case of Galactic
foregrounds or $\sigma$-clipping thresholds in the case of
extragalactic foregrounds to exclude high signal pixels (maps
expressed in terms of antenna temperature in mK units).
 }\label{tab:app:c:two}
 \end{table*}
 } 

 \newcommand{\TABLEAPPCTHREE}{
 \begin{table*}[ht]
 \begin{center}
 \begin{tabular}{rrrrrrr}
 \SUBTABLEHEADERMOMENT{CMB + Galactic dust and synchrotron}
 1024 &      0.1613 &       2.9300 &      20.0627 &       0.1471 &       3.7471 &      26.2175\\
 512 &       0.1620 &       2.8617 &      19.5551 &       0.1486 &       3.5949 &      25.1637\\
 256 &       0.1590 &       3.0031 &      20.7307 &       0.1485 &       3.5886 &      25.2111\\
  \hline
 &&&&&&\\
 \SUBTABLEHEADERMOMENT{CMB + Galactic dust and synchrotron,  $|b|>30^\circ$ }
 1024 &       0.1029 &       0.0304 &       0.0566 &       0.0786 &       0.0676 &       0.1280\\
  512 &       0.1012 &       0.0331 &       0.0437 &       0.0787 &       0.0677 &       0.1107\\
  256 &       0.0962 &       0.0400 &       0.0518 &       0.0782 &       0.0692 &       0.1128\\
  \hline
 &&&&&&\\
 \SUBTABLEHEADERMOMENT{CMB + SZ and extragalactic sources}
 1024 &       0.3004 &     125.3678 &   27163.9629 &       0.0817 &       2.3212 &      55.4750\\
 512 &       0.2290 &      87.9838 &   15230.6826  &       0.0823 &       2.2019 &      51.8950\\
 256 &       0.1457 &      28.4538 &    2166.2129  &       0.0815 &       2.0211 &      45.2755\\
 \hline
 &&&&&&\\
 \SUBTABLEHEADERMOMENT{CMB + SZ and extragalactic sources, $5 \sigma\mathrm{-clipping}$}
 1024 &       0.1090 &       0.1633 &       0.5535 &       0.0767 &       0.1281 &       0.3061\\
  512 &       0.1048 &       0.1917 &       0.4495 &       0.0773 &       0.1258 &       0.2467\\
  256 &       0.0982 &       0.1912 &       0.4427 &       0.0768 &       0.1248 &       0.2411\\
 \hline
 &&&&&&\\
 \SUBTABLEHEADERMOMENT{CMB + all foregrounds}
 1024 &       0.3257 &      98.6301 &   19666.8828 &       0.1508 &       3.8517 &      28.4794\\
  512 &       0.2628 &      58.8482 &    8779.2510 &       0.1525 &       3.6713 &      26.9822\\
  256 &       0.1946 &      13.5666 &     688.5937 &       0.1523 &       3.6406 &      26.4542\\
   \hline
   &&&&&&\\
 \SUBTABLEHEADERMOMENT{CMB + all foregrounds, $5 \sigma\mathrm{-clipping}$ and $|b|>30^\circ$}
 1024 &       0.1128 &       0.2045 &       0.6126 &       0.0812 &       0.1791 &       0.4069\\
  512 &       0.1081 &       0.2308 &       0.5769 &       0.0814 &       0.1781 &       0.3875\\
  256 &       0.1016 &       0.2304 &       0.5608 &       0.0809 &       0.1808 &       0.3966\\
  \hline
 \end{tabular}
 \end{center}
 \caption{
Statistical moments (rms, skewness and kurtosis indices) of
compositions of simulated maps. Several ``masks'' are considered:
simple Galactic cuts and/or $\sigma$-clipping thresholds (maps
expressed in terms of antenna temperature in mK units).
 }\label{tab:app:c:three}
 \end{table*}
 } 

\begin{document}


 \title{The effect of signal digitisation in CMB experiments}

\author{
  M.~Maris\inst{1}
  \and
  D.~Maino\inst{2}
  \and
  C.~Burigana\inst{3}
  \and
  A.~Mennella\inst{4}
  \and
  M.~Bersanelli\inst{2,4}
  \and
  F.~Pasian\inst{1}}

\offprints{Michele Maris (maris@ts.astro.it)}

\institute{
 INAF/Osservatorio Astronomico di Trieste, Via G.~B.~Tiepolo 11,
 I-34131, Trieste, Italy
 \and
 Dipartimento di Fisica, Universit\`a degli Studi di Milano, Via Celoria 16,
 I-20131, Milano, Italy
 \and
 IASF/CNR, Sezione di Bologna, Via Gobetti 101, I-40129, Bologna, Italy
 \and
 IASF/CNR, Sezione di Milano, Via Bassini 15, I-20131, Milano, Italy  }

   \date{Submitted to A\&A April 1, 2003}

 \abstract{
Signal digitisation may produce significant effects in balloon -
borne or space CMB experiments, when the limited bandwidth for
downlink of data requires for loss-less data compression. In fact,
the data compressibilty depends on the quantisation step $q$
applied on board by the instrument acquisition chain. In this
paper we present a study of the impact of the quantization error
in CMB experiments using, as a working case, simulated data from
the {\sc Planck}/LFI 30 and 100~GHz channels.
At TOD level, the effect of the quantization can be approximated
as a source of nearly normally distributed noise, with RMS $\simeq
q/\sqrt{12 N_{\mathrm{s}}}$, with deviations from normality
becoming relevant for a relatively small number of repeated
measures $N_{\mathrm{s}} \lsim 20$. At map level, the data
quantization alters the noise distribution and the expectation of
some higher order moments. We find a constant ratio, $\simeq
1/({\sqrt{12}\sigma/q})$, between the RMS of the quantization
noise and RMS of the instrumental noise, $\sigma$\ over the map
($\simeq 0.14$ for $\sigma/q \simeq 2$),
while, for $\sigma/q \sim 2$, the bias on the expectation for
higher order moments is comparable to their sampling variances.
Finally, we find that the quantization introduces a power excess,
$C_\ell^{ex}$, that, although related to the instrument and
mission parameters, is weakly dependent on the multipole $\ell$ at
middle and large $\ell$ and can be quite accurately subtracted.
For $\sigma/q \simeq 2$, the residual uncertainty, $\Delta
C_\ell^{ex}$, implied by this subtraction is of only
$\simeq$1--2$\%$ of the RMS uncertainty, $\Delta C_\ell^{noise}$,
on $C_\ell^{sky}$ reconstruction due to the noise power,
$C_\ell^{noise}$. Only for $\ell \lsim 30$ the quantization
removal is less accurate; in fact, the $1/f$ noise features,
although efficiently removed, increase $C_\ell^{noise}$, $\Delta
C_\ell^{noise}$, $C_\ell^{ex}$ and then $\Delta C_\ell^{ex}$;
anyway, at low multipoles $C_\ell^{\rm sky} \gg \Delta C_\ell^{\rm
noise} > \Delta C_\ell^{ex}$.
This work is based on {\sc Planck} LFI activities.

   }

 \authorrunning{M.~Maris~\etal}

 \titlerunning{Digitisation error in CMB Experiments}

 \maketitle

 \section{Introduction}
 \label{sec:introduction}

Since the successful detection of CMB anisotropy at angular scales
$\theta\gsim 7^\circ$ by the $COBE$-DMR experiment (Smoot
\etal~1992; Bennet \etal 1996; G\'orski \etal~1996), a large
number of detections on smaller angular scales with increased
sensitivity by both ground-based and balloon-borne experiments
have followed (e.g. Lasenby \etal~1998, De~Bernardis \& Masi~1998,
Bersanelli \& Maino \& Mennella~2002 for reviews). Recent exciting
results from BOOMERanG (De~Bernardis \etal~2000), MAXIMA-1 (Balbi
\etal~2000), DASI (Pryke \etal~2001) and CBI experiments (Mason
\etal~2001) have provided strong support to the inflationary
scenario for structure formation and a universe with $\Omega_0
\sim 1$.
However to fully extract the wealth of cosmological information
encoded in the CMB angular power spectrum, extreme mapping
sensitivity is required.
Only space missions can provide the proper combination of full-sky
view and low level of systematic effects that is necessary for
``precision'' CMB measurements (Danese \etal~1996), as
demonstrated by the excellent results from the NASA satellite
MAP~\footnote{http://map.gsfc.nasa.gov/} (Bennett \etal~2003).
 %
The ESA satellite {\sc Planck}~\footnote{http://planck.esa.nl}
(Mandolesi \etal~1998; Puget \etal~1998) is designed to measure the
CMB anisotropy with an accuracy
mainly set by astrophysical limits on a wide frequency range
(30--900~GHz) and with unprecedent angular resolution and sensitivity.

While space experiments benefit from unique environmental
conditions, numerous technical issues need to be addressed. One of
them is the limited bandwidth for downloading of data to the Earth
which often imposes strong limits on the available telemetry rate.
For this reason it is often necessary to use loss less compression
methods (Maris~\etal~2000a) in order to reduce the data rate.
Because the efficiency compression algorithms depends on the
quantization step applied to the observed signal by the instrument
acquisition chain on board the satellite, it is necessary to find
the optimal trade-off between the required compression rate and
the quantisation step, which needs not to be too high to preserve
the scientific quality of the measured data.

If the quantization step, $q$, is significantly smaller than the
signal root mean square (RMS), $\sigma_{\mathrm{T}}$, and no
saturation of the quantization levels occurs then the distortion
induced by the signal quantization can be well approximated by an
additive source of non-Gaussian white noise with variance $\approx
q^2/12$ (Kollar, 1994). Corrections to this simplified model are
required when $\sigma_{\mathrm{T}}/q \lsim 1$, which depend on the
kind of the total signal in input to the quantization stage of the
on-board acquisition chain (Kollar 1994). For example, in the case
of a pure white noise signal they are exponentially damped by a
factor $\exp[-2(\pi\sigma_{\mathrm{T}}/q)^2]$. On the contrary, in
the case of a sinusoidal signal (e.g. like the CMB dipole),
corrections may by quite large and no simple analytical
expressions can be written.

It is worth noting that, in general, the observing strategy
implies repeated observations of the same region of the sky on
different time scales with detectors operating at different
frequencies. Therefore, considering that a sample is the result of
the average of $N$ repeated observations, we have that for a
normally distributed signal with $\sigma_{\mathrm{T}}/q \gg 1$ the
RMS of the quantization noise, $\sigma_q$ is:

 \begin{equation}\label{eq:qnoise:general}
   \sigma_q \approx \frac{q}{\sqrt{12 N}}\, .
 \end{equation}

In CMB experiments, however, the approximations of normal
distributed signal and low $\sigma / q$ are often not strictly
applicable.
   In fact the astrophysical signal is in general not normally
   distributed, and
   it is often required to use large quantization steps (i.e.
$\sigma_{\mathrm{T}}/q \sim 1$).
   The instrumental noise power spectrum usually shows a $1/f^{\alpha}$ spectral
behaviour, with $\alpha \sim 1 - 2$ (see Burigana et al.~1997a and
Seiffert et al.~2002),
while the number of samples is not uniformly distributed when
projected onto a sky map.
 Therefore it is apparent that the validity of
eq.~(\ref{eq:qnoise:general}) needs to be assessed in the context
of space CMB experiments in order to evaluate possible deviations
from this simple formula.

In this paper we evaluate the impact of the quantization error on
CMB measurements from space and provide tools and methods to
optimize the quantization step to reach the required compression
rate with minimal impact on the scientific output.

We show examples relative to the {\sc Planck}/LFI 30 and 100~GHz
channels. However, the formalism, tools and methods developed here
are general and applicable to CMB anisotropy experiment based on
differential receivers, as well as to other kinds of experiments
with large quantization steps.

Using Monte Carlo simulations we estimate the effect of software
quantization on the data considering various quantization
functions and for different values of the quantization step. We
also discuss the impact of different experimental conditions like
the total number of repeated measures for sample, $N_s$, the white
noise RMS, $\sigma$, the $1/f^\alpha$ noise and the scanning
strategy. The impact of quantization is evaluated at three
different levels: on time ordered data, on sky maps and their
statistical moments, and on the angular power spectrum.

In Sect.~\ref{sec:analytical} we describe the analytical model of
signal quantization and discuss the adopted numerical approach.
Sect.~\ref{sec:results} describes our results in terms of effects
on TOD and maps including their statistical properties, and on the
angular power spectrum. In Sect.~\ref{sec:QRA:removal} we present
the possibility to reduce the effect of  quantization in the
recovered power spectrum. Finally, we discuss our main results and
draw the conclusions in section~\ref{sec:conclusions}. To improve
readability, some mathematical details are reported in
Appendix~\ref{appendix:alphan} while in
Appendix~\ref{appendix:ADC:quantization} we briefly verify that
the extra-quantitazion step introduced via hardware by the analog
to digital converter (ADC) quantization represents a minor effect
with respect to the subsequent software quantization extensively
discussed in the paper. In Appendix~\ref{appendix:mom_maps} we
report on statistical moments of the map adopted in the
simulations and of several foregrounds, for comparison.

\section{Analytical model and numerical simulations} \label{sec:analytical}

\subsection{Analytical model}\label{sub:sec:acq:rec:mod}

During a typical CMB experiment from space, the measured signal is
first digitized and compressed on board and subsequently
reconstructed on ground. Furthermore the scanning strategy usually
involves multiple measurements from the same pointing direction
that are subsequently averaged during ground data analysis.
Therefore in order to assess the impact of the quantization error
we need to consider the complete process, from quantization, to
reconstruction and averaging. In the following of this paper this
process will be referred to as ``QRA process''.

Let us neglect the data compression step (that is out of the scope
of our paper) and consider first the on-board quantization. In
general the instrument output will be constituted by a Time
Ordered Data stream (TOD) representing the sky temperature
anisotropy; in the case of differential receivers, for example,
the TOD will contain consecutive samples of temperature
differences (sky minus reference signal) so that the $i^{th}$\
sample, $\Delta T_i$, will be quantized as (Maris \etal\
\cite{Maris:etal:2000a}):

 \begin{equation}\label{eq:generical:quantization}
    \QDT{i} = \QUANTIZE\left[
                   \frac{
                      \DeltaTi - \DeltaTstar
                   }{ q
                   }
                  \right] \, ,
 \end{equation}

\noindent
where $\QUANTIZE[x]$\ and $\DeltaTstar$\
represent the quantization operator and a zero level.

The second step is the data reconstruction, i.e. the inverse of
eq.~(\ref{eq:generical:quantization}). This second step is
necessary because both the quantization step $q$\ and the offset
\DeltaTstar\ will not necessarily be constant during the mission;
this implies that eq.~(\ref{eq:generical:quantization}) needs to
be inverted in order to be able to compare measurements obtained
in different periods of observation. The data reconstruction step
can be defined as:

\begin{equation}\label{eq:generical:restoring}
    \RDT{i} = \left( \QDT{i}  + \OffsetQ \right) \cdot q +
    \DeltaTstar
\end{equation}

 \noindent
where \OffsetQ\ is an offset that depends on the considered
quantization operator, and is defined by:

\begin{equation}\label{eq:def:offset}
    \OffsetQ = \EXPCT[\DeltaTi]/q - \EXPCT[\QDT{i} ] \end{equation}

 \noindent
where $\EXPCT[X]$ represents the expectation of the random
variable $X$.

The value of \OffsetQ\ is dependent on the quantization operator
and assumes the values of $0$, $+1/2$, $-1/2$ for the $\floor(x)$,
$\ceil(x)$ or $\round(x)$ operators, respectively. In the case of
the $\trunc(x)$ operator the offset depends on the statistics of
the random variable $X$; if $X > 0$\ then $\trunc(x) \equiv
\floor(x)$, if $X < 0$, $\trunc(x) \equiv \ceil(x)$.

In our analytical treatment we will neglect the presence of an
offset; its effect will be discussed in the next section. Other
simplifying assumptions are that the signal is quantized in
equally spaced steps and that $\DeltaTstar$\ is always $> 0$. With
simple considerations we can estimate the effect of the QRA
process on the reconstructed maps and power spectra. We consider
the case of typical CMB anisotropy experiments consisting of
collections of many TODs each one based on the averaging over \Ns\
repetitions of the observation of the same {\it stripe} in the
sky~\footnote{For example, {\sc Planck} TODs consist of about $1.1
\times 10^4$ {\it scan} circles (for about 14 months of
observations) each of them observed 60 times. See also
section~2.2.}.

Assuming that the QRA acts mainly as an added white noise source,
the effect of quantization on the power spectrum can be easily
estimated. Let $\Clwn{,\Ns}$ the white-noise induced excess of
power and $\ClQRA{,\Ns}$\ the equivalent excess of power induced
by the QRA process; then we have that $\Clwn{,\Ns} =
\Kwn\,\sigma^2/\Ns$\ and $\ClQRA{,\Ns} = \Kqra\,q^2/(12\,\Ns)$,
where \Ns\ represents the number of times that each pixel in the
sky is measured during a single {\it stripe} and the pixel
dependent normalization constants $\Kwn$\ and $\Kqra$\ are
determined by the scanning strategy (and the processing of the
data, assumed to be a linear process). Since both quantized and
not-quantized signals are sampled and processed in the same way,
in principle we expect $\Kwn = \Kqra$. Therefore, for any $\Ns$ we
have:

 \begin{equation}\label{eq:clwn:over:clqra}
  \frac{\Clwn{,N}}{\ClQRA{,N}} =
      \left(\sqrt{12} \,  \frac{\sigma}{q}\right)^{2} \, .
 \end{equation}

 \noindent
The total noise power spectrum after quantization can be well
approximated by:

 \begin{equation}\label{eq:cl:total:noise}
  C_{\ell,N}^{total} = \Clwn{,N} \left[ 1 +
     \left(\sqrt{12} \, \frac{\sigma}{q}\right)^{-2}
     \right] \, .
 \end{equation}

 \noindent
For $\sigma/q \simeq  2$\ (as in the case of the {\sc Planck}
baseline) the noise power excess will be increased by $\simeq
2$\%. A similar relation holds for the ratio between the variance
of the quantization noise and the variance of the white noise in a
map:

 \begin{equation}\label{eq:map:total:noise}
   \sigma_{\mathrm{map},\, \mathrm{total}}^2 =
   \sigma_{\mathrm{map},\, \mathrm{wn}}^2
   \left[
     1 + \left( \sqrt{12} \frac{\sigma}{q} \right)^{-2}
   \right],
 \end{equation}

 \noindent
where $\sigma_{\mathrm{map},\, \mathrm{wn}}$\ is the RMS white
noise in the map introduced by the instrument. This formula holds
both for the noise average variance on the map and for the noise
variance of a given pixel in the map.

 \subsection{Numerical simulations}
 \label{sec:Num:Sim}

The analytical model described in the previous section allows a
simple, first-order estimation of the QRA effect. We have
performed numerical simulations to verify and quantify the
validity and the accuracy of the analytical model and to study
effects (like the presence of the skysignal and $1/f^\alpha$
noise) not accounted for in the analytical approach. The
simulations are representative of the {\sc Planck} mission. In
particular we consider here the 30~GHz and 100~GHz channels of
{\sc Planck}/LFI assuming angular resolutions of $\simeq 33'$ and
$10'$, respectively.

 \subsubsection{Simulation tools}
 \label{sec:tools}

To evaluate the effect of QRA process through numerical
simulations we have $a)$ simulated the data acquisition during the
space mission (with and without signal QRA processing) producing
TODs; $b)$ produced reconstructed sky maps and power spectra from
simulated TODs; $c)$ compared the results when including or not
the QRA process.

The simulation of the data acquisition during the mission has been
performed using the {\sc Planck} Flight Simulator developed by
Burigana et al. (\cite{Burigana:destriping}). In this code we have
included numerical tools to simulate data acquisition, signal
quantization and signal reconstruction to allow the parallel
generation of a quantized data stream plus a data stream of
reconstructed data and to perform the statistical evaluation of
the quantization error at TOD level (Maris \etal\
\cite{Maris:etal:2000b}).

The Flight Simulator is a code that simulates the CMB measurement
according to the {\sc Planck} satellite scanning strategy. {\sc
Planck} is a spinner (with the telescope observing axis nearly
perpendicular to its spin axis) that will map the sky through a
large set of near great circles at a spin rate of 1 r.p.m. The
satellite will orbit around the Lagrangian point L2 of the
Sun-Earth system maintaining the spin-axis always in the anti-Sun
direction. The measurement strategy is to acquire data from the
same sky circle for about one hour before repointing the spin axis
by approximately $2.5'$, so that the same sky pixel is measured
$\sim 60$ times in each scan circle.

The main simplifying assumptions concerning the scanning strategy
that we have considered in our simulations are: (i) no drift or
variation in the spin rate (equal to 1 r.p.m. for all the
simulation), (ii) instantaneous repointing, (iii) perfect pointing
with $85^\circ$ observing angle with respect to the spin axis.

The only astrophysical signal considered in the simulation has
been a map describing the CMB fluctuations, therefore neglecting
the effect of the dipole and of the galactic and extragalactic
contributions. To the astrophysical signal we have also added
white noise representative of the 30~GHz and 100~GHz radiometers
(with an RMS of 1.056~mK and 3.325~mK, respectively, corresponding
to integration times of 0.03~s and 0.009~s). In some simulations
we also added a 1/$f$ component to study the effect of correlated
noise on quantization; to this aim we considered two cases, with a
1/$f$ component having a knee frequency of 0.1 and 0.5 Hz,
respectively. The first is representative of a typical worst case
expected in LFI radiometers (for which we require a knee frequency
$\leq 0.05$ Hz) while the second has been studied to enhance any
possible effect connected to 1/$f$ noise but is clearly non
representative of expected performances.

The code developed by Burigana \etal\ (\cite{Burigana:destriping})
and Maino \etal\ (\cite{Maino:destriping}) has been used to
generate maps from TODs. With this code it is also possible to
significantly reduce correlated systematic effects as the
$1/f^\alpha$ noise stripes from the data before generating the
final maps. Power spectra have been obtained from the generated
maps using the {\tt anafast} code of the {\tt HEALPix} package
from G\'orsky et al. 1998, adopted here to pixelize
the sky.

 \subsubsection{Numerical procedures}
 \label{sec:procedures}

Let us now consider the time ordered data stream produced by the
Flight Simulator as an ordered sequence of measured temperature
differences $\Delta T_i$, as in eq.~(2). If we now take into
account that for each pointing direction in the sky, indicated by
the subscript $p$, on a given {\it stripe} (i.e., a scan circle
for {\sc Planck}) there will be \Ns\ repeated measurements then we
can reorder the TOD according to two indices, $(s,p)$, the first
one ($s=1, \dots, \Ns$) representing the repeated pixel
measurements and the second one representing the pointing
direction. Therefore we will have that $\Delta
T_i\equiv\DeltaT_{s,p}$ ($i \leftrightarrow (s,p)$).

Now for a single sample $(s,p)$ the quantization error introduced
by a quantization step, $q$, can be defined as:

 \begin{equation}\label{eq:quantization:error:T}
  \DeltaTE{s,p} = \DeltaTR{s,p} - \DeltaT_{s,p} \, ,
 \end{equation}

 \noindent
where the superscript $R$ indicates a sample that has been
quantized and then reconstructed. In a similar way we can define
the following quantities averaged over \Ns\ multiple measurements
for each direction, $p$, in the sky:

 \begin{equation}\label{eq:stat:idx:mean:DeltaT}
 \DeltaT_{p} = \frac{1}{\Ns} \Sigma_{s} \Delta T_{s,p},
 \end{equation}

 \begin{equation}\label{eq:stat:idx:mean:DeltaTR}
 \DeltaT_{p}^R = \frac{1}{\Ns}  \Sigma_{s} \RDT{s,p},
 \end{equation}

 \begin{equation}\label{eq:stat:idx:mean:QE}
 \DeltaT_{p}^E = \DeltaT_{p}^R - \DeltaT_{p}
            = \frac{1}{\Ns}  \Sigma_{s} \EDT{s,p} \, .
 \end{equation}

To characterize the quantization error for the LFI signal at TOD
level we study the behavior of the distribution moments of
$\DeltaT_{p}$, $\DeltaT^R_{p}$ and $\DeltaT_{p}^E$. In particular
we analyze their expectation, RMS, skewness and kurtosis as a
function of the quantization step $q$, of the quantization
function $\QUANTIZE[\cdot]$\ and of the number of samples \Ns.

The time ordered data with and without quantization have then be
used to produce maps and power spectra in order to study the
quantization error at these other two levels.

The simulations have been carried out considering three distinct
quantization steps, i.e.
   $q \approx 0.5$, 1.0, 2.0~mK
for the 30~GHz channel and
  $q \approx 1.2$, 2.4, 4.9~mK
for the 100~GHz channel. These three values of $q$ correspond, for
each frequency channel, to $\sigma/q \simeq 2$, 1 and 1/2,
respectively. The case $\sigma/q \simeq 2$ corresponds to the
current {\sc Planck} baseline; the case $\sigma/q \sim 1/2$ is
representative of a case in which the noise model for the
quantization distortion is expected to fail, while the
intermediate case was chosen evaluate possible deviations of the
dependence of the quantization error on $\sigma/q$ from a power
law.

 \FIGUREONE

 \section{Results}
 \label{sec:results}

In this section we discuss the effect of signal quantization at
the three different levels of time ordered data, final maps and
power spectra. The results presented in this section have been
produced considering the quantization operator $\floor(x)$; this
assumption can be made without any loss of generality, as all the
quantization operators can be considered equivalent provided that
the proper offset is applied in the reconstruction formula,
eq.~\ref{eq:generical:restoring}.

 \subsection{Effect of quantization at TOD level}
 \label{sub:sec:results:bias}

To study the effect of quantization at the level of the
time-ordered data stream we focus on the statistical properties of
the data-stream by evaluating the various moments of the
statistical distribution of the quantization error.

 \subsubsection{Mean value of the quantization error distribution}
 \label{sub:sec:error_mean_value}

The first statistical quantity that we have analyzed is the mean
value of the quantization error to evaluate if the QRA process
introduces any spurious bias in the data and how this varies with
$q$. This is relevant in the context of space missions, in which
possible variations in the noise properties of the detectors will
require slight modifications in time of $q$ in order to maintain
constant the $\sigma/q$ ratio required for optimal compression.
Therefore if any bias in the data is introduced by quantization it
is reasonable to expect that this bias will vary in time (because
of the changes in $q$) leading to systematic errors in the final
maps and power spectra.

We have calculated the mean value of the quantization error
($\DeltaTE{60} = \mean[\DeltaTE{60,p}]$, see
eq.~\ref{eq:stat:idx:mean:QE}) for a $\sim$ 1 year-long data
stream, considering different values of the quantization step $q$.
Our results show that quantization introduces a bias in the data
at a level $|\DeltaTE{60}|/q < 10^{-3}$ which implies
$|\DeltaTE{60}| < 2$~$\mu$K even for the largest values of $q$.
This offset effect is small, and in addition it can be efficiently
removed by destriping algorithms.

Therefore we can expect systematic effects in the time ordered
data at a very low level and the time scales of the updating of
$q$ on board ($>$ 24~hours). Furthermore these effects can be
efficiently reduced by applying destriping and/or map-making
methods to the time-ordered data and therefore they do not
represent a concern\footnote{In fact, the current destriping codes
are able to significantly reduce the impact of much larger drifts,
such as those induced by the $1/f^\alpha$ noise or by periodic
fluctuations induced by thermal fluctuations.}.

 \subsubsection{RMS of the quantization error distribution}
 \label{sub:sec:error_rms}

The rms of the quatization error can be easily evaluated from the
theoretical eq.~(\ref{eq:qnoise:general}). The validity of this
relationship has been checked numerically by calculating the rms
of the quantization error (indicated by $S_N$) for several TODs
generated with various values of $q$ and $N$ around the reference
values $N_{\mathrm{0}} = 60$ and $\qZero = 0.305$~mK/adu,
equivalent to $\sigma/q = 3.462$\ for the 30 GHz channel and
$\sigma/q = 10.902$\ for the 100 GHz channel. A fit of $S_N$ with
the relation:

 \begin{equation}
 \ln S_{N} =
 \ln\Sigma_{0}
 + m_{\mathrm{N}}
 \ln(\Ns /N_{\mathrm{0}})
 +
 m_{\mathrm{q}} \ln( q/\qZero ) \, .
 \end{equation}

 \noindent
yielded $\Sigma_{0} \approx \qZero / \sqrt{12 N_0}$,
$m_{\mathrm{N}} \approx -1/2$ and $m_{\mathrm{q}} \approx 1$ with
an accuracy up to the third decimal figure. This confirms that in
our case eq.~(\ref{eq:qnoise:general}) is a very good
approximation of the rms value of the quantization error
distribution.


 \FIGURETWO

 \subsubsection{Higher moments of the quantization error distribution}
 \label{sub:higher_moments}


In the ideal case of a normally distributed random signal with
$\sigma /q \gg 1$ the QRA error will also be nearly normally
distributed and the indices of skewness and kurtosis~\footnote{In
what follows we will use sometimes the terms ``skewness'' and
``kurtosis'' instead of skewness and kurtosis index. These
indices, combinations of the distribution statistical moments
$\mu_i$, are defined by $\mathrm{Skewness \, index} =
\mu_3/\mu_2^{3/2}$ snd $\mathrm{Kurtosis \, index} =
\mu_4/\mu_2^{2}-3$.} are expected to vanish. On the other hand if
$\sigma / q$ approaches 1 then it is reasonable to expect a change
in the values of the higher moments of the statistical
distribution for the QRA error.

From theory, the effect on higher moments may be evaluated from
the characteristic function for the distribution of the sampled
values after QRA processing $\CharFuncQRAData(\omega) =
\CharFuncQRA(\omega) \CharFuncData(\omega)$, where $\omega$\ is a
real variable, $\CharFuncQRA(\omega)$\ is the characteristic
function for the QRA process (see appendix~\ref{appendix:alphan}),
$\CharFuncData(\omega)$\ is the corresponding function for the
averaged input data. The expectation for central moments of order
$k\ge0$ being $\mu_k = (-i)^k d^k \CharFuncQRAData(\omega) /
d\omega^k |_{\omega = 0}$.
Since the QRA error is even distributed around zero expectations
for its moments for $k$ odd are null. If the same is true for the
distribution of the unprocessed data then the expectations for all
the $\mu_k$\ with $k$ odd is null.
The same does not hold for the expectation of moments with $k$
even, in this case the QRA process may introduce a not null bias
for some or all of them.
In addition the QRA process randomly scatters about their
expectation both odd and even moments.
Then QRA process randomly changes the measured value of moments of
any order $k$ for a given realization respect to the values which
have been measured in the case of no-QRA.
Higher order moments, at least up to the fourth order, are needed
to full-characterize the statistical properties of the QRA error.
However the complete study of these features is beyond the scope
of this work and we limit our investigation to few cases
illustrative of the methods which may be applied to study them and
the kind of perturbation which may be expected on real data. We
may attempt a deeper investigation of this effect as numerical
tools will improve enough to allow the generation and the analysis
of a large data set of simulated missions.

 %
 %


To evaluate the departure of the averaged quantization error from
a normal distribution we have calculated the skewness and the
kurtosis for the distribution of $\DeltaT_{p}$, $\DeltaT_{p}^R$
and $\DeltaT_{p}^E$ for various values of $q$\ and $\Ns$.

 \FIGURETWOBIS

As expected, the QRA process alters the sampling skewness in the
input signal, so that it has meaning to put:

{\small
 $$
  \frac{\Delta \skewness} {\skewness[\DeltaT_{p}]} =
   \frac{(\skewness[\DeltaT_{p}^R] - \skewness[\DeltaT_{p}])}
{\skewness[\DeltaT_{p}]} \neq 0 \, .
 $$
}

 \noindent
In our simulation we observe values of $|\Delta \skewness /
\skewness[\DeltaT_{p}]|$\ up to $50\%$ at the TOD level,
comparable to the random fluctuations induced by the sampling
variance. It is worth noting that this value depends on $N_s$ so
that the variation in skewness decreases with increasing the
number of samples of the same data point and that on the final map
(in which each pixel will be the result of averaging a large
number of measurement, especially close to the ecliptic poles) the
impact on the skewness of the CMB distribution is even smaller
(see Table~\ref{tab:qerror:mapstat:nongaus}) and varying with the
colatitude.

In Fig.~\ref{fig:tod:kurtosis} we show the behavior of the
kurtosis for the QRA error
($\EXPCT[\kurtosis[\DeltaT_{p}^{\mathrm{E}}]]$) as a function of
\Ns. From the plot it is apparent that the expectation of the
kurtosis goes to zero for increasing values of \Ns. For an input
signal with small expected kurtosis, the value of the sampling
kurtosis at TOD level can be changed by the QRA process by a
factor up to $\sim 2$, again comparable with the sampling variance
for the kurtosis in our TODs. As for the skewness the impact on
the kurtosis on maps is even smaller. However it is important to
recall that, while skewness is changed randomly by the QLA, i.e.
no bias is added by QRA to the skewness estimator, the same does
not hold for the kurtosis. For the kurtosis, the bias is
$-3/[1+12(\sigma/q)^2]^2$, equivalent to $\simeq -0.001$ for
$\sigma/q \simeq 2$. However, once the quantization step and the
scanning strategy is defined, it is a simple matter to calculate
this bias and to remove it.

 \FIGURETHREE

A non-zero value of the kurtosis in the distribution of the QRA
error changes the statistical significance of the confidence
limits respect to the usual definition of ``standard error'' for
which ``$1\sigma$\ corresponds to a 68.27\% confidence level''. To
recover the standard error definition we have scaled the variance
of the quantization error by a multiplicative factor
$\alpha_{N_{\mathrm{s}}} \gsim 1$, which decreases monotonically
for increasing \Ns, whose $\lim_{N_{\mathrm{s}} \rightarrow
\infty} \alpha_{N_{\mathrm{s}}} = 1$, and $\alpha_1 = 1.18$.
Appendix \ref{appendix:alphan} describes the computation of
$\alpha_{N_{\mathrm{s}}}$.

These results allow to safely apply the usual error propagation to
combine both signal plus noise and the QRA error with an accuracy
better than some $\mu$K. The $1\sigma$ confidence level for a
given temperature average, $\mu_{N_{\mathrm{s}}}$ may be expressed
as:

 \begin{equation}\label{eq:alpha:1sigma:noise}
  \mu_{N_{\mathrm{s}}} \pm \sigma_{N_{\mathrm{s}}} \pm
\alpha_{N_{\mathrm{s}}}
  \sigma_{q,N_{\mathrm{s}}}
  \approx
   \mu_{N_{\mathrm{s}}}
   \pm
   \sqrt{\sigma_{N_{\mathrm{s}}}^{2}+ \alpha_{N_{\mathrm{s}}}^2
\sigma_{q,N_{\mathrm{s}}}^{2}} \, .
 \end{equation}

 \noindent
The first term down the square root represents the effect of the
white-noise, while the second term the effect of the QRA error.
Note that since the $\alpha_{N}$\ correction equations
(\ref{eq:cl:total:noise}) and (\ref{eq:map:total:noise}) are valid
only for \Ns\ sufficiently large to have $|\alpha_{N_{\mathrm{s}}}
- 1| \ll 1$. Otherwise the $\left( \sqrt{12} {\sigma}/{q}
\right)^{-2}$\ has to be scaled by $\alpha_{N_{\mathrm{s}}}^2$\ in
eq.~(\ref{eq:cl:total:noise}) and eq.~(\ref{eq:map:total:noise}).
In the case of {\sc Planck}-LFI the measurement redundancy
($\Ns\sim 60$) is such that the distributions of the QRA error is
sufficiently well approximated by a normal distribution to apply
the usual formula for the error propagation.

 \FIGUREFOUR

\subsection{Effect on the reconstructed map}
\label{sec:pw:sp}

 \TABLEONE
 \TABLEONEB


In Fig.~\ref{fig:MAP:qerror}\ (top panel) we show a map of the QRA
error for $\sigma/q \approx 2$ for a 30~GHz detector. The map is
pixelised according to the HEALPix scheme (G\`orski et al. 1998)
with a pixel-size of about 13.7~arcmin corresponding to
$N_{\mathrm{side}}$ ($12\times N_{\mathrm{side}}^2$ is the number
of pixels in the map). To generate the map we first produced two
TODs of the same simulation (including CMB, white and 1/$f$ noise)
one with quantization and the other without. The error map was
then obtained by differencing the maps generated by the two data
streams. A simple ``visual'' inspection of the map indicates the
absence of evident correlate structures\footnote{The darker narrow
band visible in the map is not due to quantization but to the
overlapping of a subset of scan circles. In this region the
integration time is greater and, therefore, the noise level is
smaller.} at few $\mu$K level. Although the peak-to-peak variation
of the quantization error on the map may appear large
($\sim$50~$\mu$K) we must underline that quantization does not act
as a correlated systematic effect, but is mainly an added white
noise at a few \% level over the detector white noise (see
eq.~(\ref{eq:map:total:noise})).

In the bottom panel of Fig.~\ref{fig:MAP:qerror} we show the ratio
between the QRA error distribution and a normal distribution with
the same mean and RMS, indicating that the QRA error is not
completely normal distributed, with a number of error samples
larger than $2\sigma$ that is less compared to a normally
distributed noise. These ``non-Gaussian samples'', however, are
mostly contained in a symmetric band at ecliptic latitudes between
$\pm 15^\circ$, and owing to the larger number of samples, this
feature of the histogram is less large for the 100~GHz channel
than the 30~GHz.


In Tab.~\ref{tab:qerror:mapstat} we report the \% noise increase
determined by quantization for various values of $\sigma / q$. The
comparison between the values determined by simulations with those
calculated by eq.~(\ref{eq:map:total:noise}) shows the departure
from the simple theoretical model described in
Sect.~\ref{sub:sec:acq:rec:mod}. For $\sigma / q \sim 2$
eq.~(\ref{eq:map:total:noise}) still represents a good estimate of
the main effect of quantization.



There is an interesting parallelism between what happens to
moments for a TOD of pure white noise when the QRA process is
introduced, and what happens for the same moments in the map
obtained from the same TOD. To illustrate it, let be to denote
with $\mu_{k,map,wn,QRA}$ and $\mu_{k,TOD,wn,QRA}$ the $k$-th
moment for the map obtained from QRA processed data and its
corresponding TOD from which the map is obtained, and with
$\mu_{k,map,wn}$ and $\mu_{k,TOD,wn}$\ the corresponding moments
obtained for the map without quantization and the corresponding
TOD. Looking to pixels as averages of $N_p$\ samples, where
$N_p\ge1$ is the number of TOD samples entering a given pixel $p$
in the map, it is possible to classify pixels as a function of
$N_p$. If $[N_p]$\ is the class of pixels averages of $N_p$\
samples, $g(N_p)$ the partition function for the pixels over the
various classes it is easy to demonstrate that, for any value of
\Ns\ and $\sigma/q$, and $k=2$\ the relation

 \begin{equation}\label{eq:map:tod:moments:relation}
     \frac{\mu_{k,\mathrm{map}, wn,\mathrm{QRA}}}
      {\mu_{k, \mathrm{map},wn}}
     =
     \frac{\mu_{k, \mathrm{TOD}, wn, \mathrm{QRA}}}
     {\mu_{k, \mathrm{TOD},    wn}}
 \end{equation}

 \noindent
holds exactly. From this eq.~(\ref{eq:map:total:noise}) may be
easily derived. For higher order moments ($k>2$) the relation in
eq.~(\ref{eq:map:tod:moments:relation}) holds better and better as
$\Ns$ is larger and larger. Indeed, if \Ns\ is not large enough
other terms including $g(N_p)$\ will affect the right hand side of
eq.~(\ref{eq:map:total:noise}). In the case of {\sc Planck}\
having $\Ns = 60 $ these terms are negligible at least up to $k =
4$. In conclusion, from the same reasoning leading to
eq.~(\ref{eq:map:total:noise}) it is possible to draw equations
linking the variations of higher order moments on TODs to the
corresponding variations results of higher order moments on maps.

As anticipated in the section related to the TODs, we compared the
kurtosis in a map which has been QRA processed (\KurtQRA), against
the kurtosis in the corresponding original map (\KurtOrig) giving
the ratio of the variation ($|\Delta K/K| = |\KurtQRA -
\KurtOrig|/\KurtOrig$). In general $|\Delta K|/K$\ is greater for
the 30~GHz channel than for the 100~GHz. In particular for the
30~GHz channel considering maps which does not contain $1/f$ noise
and consequently have not been destriped the kurtosis may be
varied even of a 30\% by the QRA process.
However, comparing maps which does not contain $1/f$ noise and have
not been destriped with maps originated from data containing $1/f$
noise and destriped it is possible to see that the relative effect
on the kurtosis is much smaller than in the previous case. This is
nothing else than the fact that $1/f$ noise and destriping
increases the kurtosis of the unquantized signal given at the
denominator of the $|\Delta K/K|$\ ratio, but leaves practically
unchanged the $\Delta K$ difference.
From  Tab.~\ref{tab:qerror:mapstat}) it is also possible to see
how $|\Delta K/K|$\ scales with $\sigma/q$\ both for the 30~GHz
and 100~GHz maps.
Comparing maps which contains even the $1/f$ noise and have been
destriped (last column of $|\Delta K/K| \lsim 4.2\%$ ($1.5\%$).
For both 30~GHz and 100~GHz maps $|\Delta K/K|$\ increases up to a
factor of about 7 decreasing $\sigma/q$\ from $2$ down to $1/2$.

The square root of the sampling variance of $K$ is
$\sqrt{24/\Npix}$ (see e.g. Kendal \& Stuart 1977), i.e.
$\sqrt{2}/N_{\mathrm{side}}$ in the case of the HEALPix scheme,
over a map with $\Npix$ pixels obtained from a single receiver
(radiometer for LFI). For maps at $N_{\mathrm{side}} = 256$ (512)
this gives $5.5\times10^{-3}$ ($2.8\times10^{-3}$), which is
equivalent to a sensitivity $|\Delta K/K| \simeq 30-50\%$ (20\%)
over a map from a single receiver in the case of a sky of pure CMB
fluctuations (see Tab.~C.1 in Appendix~C). When $|\Delta K/K|$\ is
scaled to take into account that the final map is derived from a
combination of more radiometers, the sensitivity to $K$ improves
to 0.0028 (0.00057) at 30~GHz (100~GHz), equivalent to $|\Delta
K/K| \simeq 15-25\%$ (4\%). The $|\Delta K/K|$ values induced by
the QRA process (see Tab.~2), although not large, is not
negligible in the case of maps of pure CMB anisotropy plus noise,
particularly because it is not a statistical but a systematic
effect. At 30~GHz, the single receiver noise is small, and the
above conclusion does not change significantly including also
noise contribution to the kurtosis, while at 100~GHz, where a
single receiver noise is quite large for LFI, the reference value
of $K$ changes significantly including or not the noise, and the
comparison between $|\Delta K/K|$ representing the sensitivity and
that reprenting the QRA effect may depend strongly on the
considered realization.

On the other hand, the microwave sky includes fluctuations
also from Galactic and extragalactic foregrounds, with
remarkable non Gaussian distributions.
In Appendix~C we report the statistical moments of
the most relevant microwave components and
of composite maps. As evident, even including Galactic cuts
and removing bright sources and Sunyaev-Zeldovich effects
(see last three lines of Tab.~C.3), foregrounds change
the sky kurtosis index with respect to the case of a pure
CMB anisotropy sky much more than the QRA process.
Therefore, the foreground removal is an aspect much more
important than QRA process for the estimation of CMB anisotropy
high order moments.



 \TABLETWO

Finally in Fig.~\ref{fig:lat:dependence} we show the variation
with the ecliptic latitude of the variance of the QRA error, which
decreases as a function of $\sin\theta$, where $\theta$ is the
pixel colatitude. The figure also shows that for $\theta >
15^\circ$ the variance can be well approximated with
$a + b \sin \theta/\sin 50^\circ$, where the best-fit
coefficients $a$ and $b$ are listed in the last two columns of
Table~\ref{tab:qerror:mapstat}. This relationship represents a
useful method to estimate the rms QRA error at a given latitude.

\subsection{Effect on the power spectrum}

In Fig.~\ref{fig:pwr:spc} we show an example relative to a {\sc
Planck}-LFI 30~GHz radiometer that highlights the qualitative
features of the quantization effect on the measured power
spectrum. To enhance the effect we have chosen a value $\sigma / q
\sim 0.5$, which is clearly not representative of typical
conditions.

The figure shows three power spectra. The first one (lower curve)
is the power spectrum of the original map. The second (middle
curve) is the power spectrum of the map ``observed'' by the
simulator without quantization; the power increase at high
multipoles is due to the instrumental white noise. The third
(upper curve) is the power spectrum of the ``observed'' map
including quantization. The inset  is an enlargement of the power
spectrum for $400 \leq \ell \leq 767$.

From the inset it is evident that the QRA process introduces a
power excess in the spectrum.
A closer look reveals that most of the features of the original
power spectrum are preserved.
At multipoles larger than a critical value~\footnote{$\ell_{\rm
crit} \sim 70$ for the 100~GHz channel and $\ell_{\rm crit} \sim
30$ for the 30~GHz channel.} ($\ell > \ell_{\rm crit})$ this power
excess approaches a value, $\ChlQRA{,60}$, nearly independent on
$\ell$ and dependent on $\Ns$ (we use here $\Ns = 60$).
The value $\ChlQRA{,60}$\ can be used as a convenient estimator of
the quantization effect on the power spectrum at large multipoles
and can be quite well estimated by the average of the power excess
for $\ell > \ell_{\rm crit}$:

 \begin{equation}
   \ChlQRA{,60} =
    \frac{
      \sum_{\ell=\ell_{\rm crit}}^{\mathrm{max}(\ell)} \ClQRA{,60}
    }{
      \mathrm{max}(\ell) - \ell_{\rm crit} + 1
    }
 \end{equation}

 \noindent
where $\mathrm{max}(\ell) \simeq 750$ and 1500 respectively for
the 30 and 100~GHz frequency channels.

The power spectra of absolute and relative quantization error
maps, in terms of both $C_{\ell}$\ and $\delta T_{\ell}$\ are
shown in Fig.~\ref{fig:pwr:excess} for different values of $\sigma
/ q$. The power excess varies roughly as a power law of
$\sigma/q$, in good agreement with eq.~(\ref{eq:cl:total:noise}).
 %
 %



Table~\ref{tab:cl70}\ gives $\ChlQRA{,60}$\ for three quantization
steps for a single radiometer at the 30 and 100~GHz frequency
channels. The simulations from which $\ChlQRA{,60}$\ are derived
assuming the nominal {\sc PLANCK} same scanning strategy and $1/f$
noise knee frequency. In addition we report two cases obtained
changing the knee frequency and the scanning strategy, which will
be discussed at the end of this section.

A fit of the data obtained by simulations with
eq.~(\ref{eq:cl:total:noise}) yielded a residual log-log
correlation coefficient $|1-r_{\mathrm{log-log}}| < 10^{-5}$,
which indicates an excellent agreement with the expected $q^2$\
dependence. This allows to parametrize~\footnote{In this work we
have not studied numerically (because of computation time limits)
the dependence of $\ClQRA{}$ on \Ns . So the $1/\Ns$\ dependence
of eq.~(\ref{eq:clwn:over:clqra}) $\ClQRA{}$\ as a function of
$\sigma$, $\sigma/q$, \Ns\ has been assumed.}:

 \begin{equation}\label{eq:ClQRA:1}
  \ClQRA{} = \ClQRA{,1} \frac{\sigma^2}{\Ns}
             \left( \frac{\sigma}{q} \right)^{-2}.
 \end{equation}

The normalization factor $\ClQRA{,1}$ can be easily determined
from the above-mentioned best-fit; a good approximation can be
obtained by replacing $\ClQRA{,1}$\ with $\Keff / \Npix$, where
$\Npix$ represents the number of pixels on the map and $\Keff$\ is
a normalization factor $0.72 < \Keff < 1$ that depends on details
such as: the scanning strategy, the pixelization scheme and the
geometry of the instrument. A further approximation $\Keff = 1$
yields:

 \begin{equation}\label{eq:ClQRA:2}
  \ClQRA{} \approx \frac{\sigma^2}{12 \Ns \Npix}
             \left( \frac{\sigma}{q} \right)^{-2},
 \end{equation}

 \noindent
which is a useful relationship to estimate upper limits to the
quantization effect on the recovered power spectrum.

Finally we can determine the ratio $K_{wn}^2/\Kqra$\ defined in
eq.~(\ref{eq:clwn:over:clqra})
by fitting $\log \ClQRA{,60}$\ versus $\log
\left(\sqrt{12}\,\sigma/q\right)^{-2}$ and taking the ratio with
the power spectra on the unquantized noise. This yields
$K_{wn}^2/\Kqra = 1.024$, that represents a good approximation of
the ideal value $K_{wn}^2/\Kqra = 1$. This is a further
confirmation that eq.~(\ref{eq:clwn:over:clqra}) and
eq.~(\ref{eq:cl:total:noise}) well describe the quantization
effect on the power spectrum.
From these simulations the total power excess due to white noise
plus the QRA processing is:

 \begin{equation}
 \langle C_{\ell}^{\mathrm{exc}} \rangle =
 \langle C_{\ell,\mathrm{WN}} \rangle
 \left[
   1 + \left(3.55 \frac{\sigma}{q}\right)^{-2}
 \right].
 \end{equation}




On the other hand, as evident in Fig.~\ref{fig:pwr:spc} referring
to realistic simulations including the $1/f$\ noise and the
application of the destriping algorithm, for $\ell < \ell_{\rm
crit}$, the power excess is no longer constant, but shows an
increase for decreasing $\ell$.
However, for $\sigma/q \approx 2$\ the deviation from a constant
power excess at low $\ell$ is not large. For example, for the
30~GHz channel it is equivalent to $\lsim 16\%$ increase of the
power excess
(50\% for the 100~GHz). For $\sigma/q \approx 1$ the deviation is
significant up to $7.5 \ClQRA{}$\ at 30~GHz and $16.8 \ClQRA{}$\
at 100~GHz. Although It is not easy to parametrize this effect,
$\ClQRA{}$\ may be well described by a second or third order
polynomial of $\log \ell$\ in the log-log space (Maris
\cite{Maris:2002a}) for $\ell > 1$, which for $\ell > 70$\ well
fits the constant power excess previously assumed. However the
results of this fit can not be easily generalized as in the
previous case, since details such as the polynomial degree are
related not only to the mission parameters but also to the
assumptions implicit into the destriping procedure and the
parameters of the $1/f$ noise. This is supported by the comparison
of simulations obtained for different values of $\fknee$ (see
Tab.~\ref{tab:cl70:fit} for the 30~GHz frequency channel). No
significant differences appear for $\ell > 30$ by changing \fknee
from 0.1~Hz to 0.5~Hz, while at $\ell < 30$\ the power excess
changes up to a factor of two.

 \TABLETHREE

A change in the scanning strategy instead, affects the power
excess at both low and large $\ell$. In Tab.~\ref{tab:cl70:fit}\
is shown what happens allowing for a precession of the spin axis
during the mission. $\ClQRA{}$\ increases by about 7.6\% at high
$\ell$, and as much as 82\% at low $\ell$. Since the same sequence
of random noise samples has been used for both the precessed and
un-precessed scanning strategy, these changes are due to the
different weighting of the samples in the sky induced by the
precession of the spin axis.

  \section{Removing the effect of the QRA error in the power spectrum estimation}\label{sec:QRA:removal}

We have seen that quantization effects on the power spectrum are
significantly smaller than the noise and that they are well
represented by few parameters. Here we discuss the possibility to
remove quantisation excess noise in the data analysis from the
final estimated power spectrum. Of course, this is of particular
interest at middle and large multipoles while at low multipoles
the noise power spectrum is much smaller. The estimation of the
sky power spectrum can be derived from the power spectrum obtained
from the observed map by subtracting the sum of the expected noise
power spectrum and of the expected QRA power spectrum:

 \newcommand{\Exp}{\mbox{$\mathrm{E}$}}

 \newcommand{\Clnoise}[1]{\mbox{$C_{\ell#1}^{noise}$}}

 \begin{equation}
     C_l^{\mathrm{sky}} = C_l^{\mathrm{Obs}} - \Exp[\Clnoise{}] - \Exp[\ClQRA{}] \, .
 \end{equation}

 \noindent
 %
The accuracy of the subtraction of the QRA effect is determined by
the accuracy of the model for \ClQRA\ and by the dispersion of the
true values of \ClQRA\ about their expectation.
The typical accuracy of this subtraction at large $\ell$, as
evaluated from the last column of Tab.~\ref{tab:cl70}, is of about
5\% (6\%) for the 100 (30)~GHz channel, leaving a residual (i.e.
after subtraction) QRA error equivalent to an unsubtracted QRA
error with a quantization step at least four times smaller than
that really used.

 %
 %

The RMS of the expected QRA power spectrum decreases with $\ell$,
analogously to the case of the RMS of white noise that exibits a
typical $1/\sqrt{\ell}$ dependence. We expect then that the
accuracy of the subtraction of the QRA effect improves with
increasing $\ell$. For example, tests with the simulated power
spectra reported in section~\ref{sec:results}\ show that the RMS
of the QRA power spectrum
ranges between
$\sim 10^{-11}$\ mK$^2$ and $\sim \mathrm{few} \times 10^{-10}$\
mK$^2$\ at $\ell \sim  1500$ (750) for the 100 (30)~GHz channel
when $\sigma/q$\ ranges between 2 and 1/2. As a consequence, the
power excess is reduced up to a factor 45 (34) for the 100
(30)~GHz channel at very large $\ell$, equivalent to an
unsubtracted QRA error with a quantization step reduced by a
factor $\sim 6-7$.
On the contrary, at $\ell$ of few tens the RMS of the expected QRA
power spectrum is of the same order of magnitude of the power
spectrum itself, $C^{\mathrm{\mathrm{exc}}}_{l,60}$.
For example at $\ell \sim 15$ it ranges between
$\mathrm{some} \times 10^{-10}$\ mK$^2$\ and $\mathrm{some} \times
10^{-8}$\ mK$^2$\ for the 100~GHz channel, and between
$\mathrm{few} \times 10^{-10}$\ mK$^2$\ and $\mathrm{few} \times
10^{-8}$\ mK$^2$\ for the 30~GHz channel. Compared with the
averaged power excess for $\ell \lsim 30$ this is equivalent to a
reduction of a power excess between $\sim 50$\% and a factor of
two.

We searched for possible improvements of the removal procedure by
with a polynomial fit of $\log C^{\mathrm{exc}}_{l,60}$ versus
$\log \ell$. We find that the best fitting relation depends on the
frequency channel, the value $\sigma/q$, the scanning strategy and
the beam position so that a proper fit for each receiver and
scanning strategy would be required.
However, our simulations show that for $\sigma/q \approx 2$ the
values of $C^{\mathrm{\mathrm{exc}}}_{l,60}$\ are very dispersed
around the mean for $\ell < 30$, while for $\ell > 70$ the
goodness of fit does not depend significantly on the degree of the
polynomial. Therefore a simple linear fit is adequate.

%

In order to evaluate the final impact of the accuracy of the
subtraction of the QRA quantization effect in the final estimation
of the sky power spectrum it is crucial to compare the RMS of the
power spectrum of QRA quantization error with the RMS of the noise
power spectrum that sets the fundamental uncertainty in the sky
power spectrum recovery.

We require that

 \begin{equation}\label{eq:fiducial}
\mathrm{RMS}[C^{\mathrm{\mathrm{exc}}}_{l,60}] <
\mathrm{RMS}[C^{\mathrm{\mathrm{noise}}}_{l}] / \lambda \, ,
 \end{equation}

 \noindent
where the constant $\lambda$ has to be set large enough to reduce
the residual QRA quantization error to an acceptable level (for
example, $\lambda > 5$ implies a residual QRA quantization error
less than 20\% of the unsubtractable instrumental noise).

After having applied the destriping algorithm, the instrumental
noise is essentially white noise dominated and we can then assume

 \begin{equation}
\mathrm{RMS}[C^{\mathrm{\mathrm{noise}}}_{l}] \simeq
\mathrm{RMS}[C^{\mathrm{\mathrm{WN}}}_{l}] \simeq
\mathrm{E}[C^{\mathrm{\mathrm{WN}}}_{l}] / \sqrt{\ell} .
 \end{equation}

As discussed before, $C^{\mathrm{exc}}_{l,60}$ depends on the
considered receiver and scanning strategy; of course, the same
holds for $\mathrm{RMS}[C^{\mathrm{\mathrm{exc}}}_{l,60}]$. It is
then impossible to rewrite in general the
condition~(\ref{eq:fiducial}) in terms of $\sigma/q$.

To circumvent this problem we have searched for a law, $U_\ell$,
which gives at each $\ell$ the maximum among the ratios
$\mathrm{RMS}[C^{\mathrm{\mathrm{exc}}}_{l,60}] /
C^{\mathrm{exc}}_{l,60}$ found for the set of simulations
considered here. We find that the approximate law

 \begin{equation}\label{eq:upplim}
U_\ell \simeq 0.93 \ell^{-0.57}
 \end{equation}

\noindent is a quite good approximation for the whole range of
multipoles relevant here.

The condition expressed by eq.~(\ref{eq:fiducial}) is then
satisfied provided that

 \begin{equation}\label{eq:fiducial1}
0.93 \ell^{-0.57} \mathrm{E}[C^{\mathrm{\mathrm{exc}}}_{l,60}]
\lsim \mathrm{E}[C^{\mathrm{\mathrm{WN}}}_{l}] / \sqrt{\ell} /
\lambda \, .
 \end{equation}

At middle and large multipoles the ratio between the variances of
the white noise and of the quantization error equals that between
the two corresponding power spectra. By using
eq.~(\ref{eq:map:total:noise}), the condition~(\ref{eq:fiducial1})
can be then easily rewritten in terms of the $\sigma/q$ ratio as:

 \begin{equation}\label{eq:fiducial2}
\left({\sigma \over q}\right)^2 \gsim {0.93 \over 12 }
\ell^{-0.07} \lambda \, .
 \end{equation}

At $\ell \gsim 30$ (500) this condition is clearly satisfied even
for $\lambda \gsim 4$, 16 or 65 (5, 20, 80) respectively for
$\sigma/q \simeq 1/2$, 1 or 2.

In conclusion, this analysis demonstrates that it is possible to
subtract the quantization error impact in the sky power spectrum
estimation from {\sc Planck} data at a level of accuracy which
makes its residual effect very small and significantly smaller
that the intrinsic unsubtractable uncertainty due to the noise.

However it should be noted that these results are obtained with
the assumption of stationary noise. In particular the application
of this correction requires to consider every changes in the
detector calibration which affects $q$\ along the mission.
Moreover, the previous result about the sensitivity of $\ClQRA{}$
on the scanning strategy imposes to consider that the final
$\ClQRA{}$\ to be subtracted from the power spectrum shall be
evaluated by numerical simulations to take into account the real
scanning strategy. This is particularly relevant for a multi-horn
instrument like {\sc Planck}, where a full evaluation of
$\ClQRA{}$\ shall be obtained only with a full simulation of the
mission with each horn observing somewhat differently the sky.

 \section{Conclusions}\label{sec:conclusions}

An unprecedented amount of high quality data are expected from the
new generation of CMB experiments designed for extended, high
resolution imaging of the microwave sky. These data sets will be
analyzed to measure the angular power spectrum to high precision,
and also to search for low amplitude statistical properties of the
CMB distribution, such as non-gaussianity or high-order
autocorrelation functions. It is therefore of interest to
accurately consider the effect of instrumental systematics that
could significantly perturb the statistical properties of the
noise and signal distribution both in the map and in the angular
power spectrum recovery.

For a space mission, the need of minimizing the telemetry rate
requires that the signal digitization in terms of the $\sigma/q$
ratio be as low as possible. On the other hand, a poor signal
sampling, combined with the adopted scanning strategy, may produce
subtle effects impacting the statistics. In this paper we have
analyzed the impact of signal digitization, reconstruction and
averaging of a full-sky CMB survey. The effect of the QRA process
has been analyzed at the level of data streams, maps and power
spectra, both by analytical approximations and numerical
simulations. We have simulated the effect for the case of the {\sc
Planck}/LFI experiment, but the methods presented here and the
main conclusions can be applied to any other experiment. The
simulated signal includes white and $1/f$-type noise and a
realization of CMB anisotropy, while the effect of QRA process on
high order statistics has been compared with that from main
Galactic and extragalactic foregrounds.

Because the signal is dominated by white noise, the $\sigma/q$\
ratio is the main parameter describing the QRA scheme. At the
level of data streams we studied the distribution of QRA errors as
a function of $\sigma/q$ and of the number $\Ns = 3$, 4, $\dots$,
60 of repeated pixel observations obtained by the PLANCK scanning
strategy. As expected, when $\Ns$ is large the QRA error is well
approximated by a normally distributed random variable, and
standard error propagation may be applied.

For large \Ns\, most of the numerical results at the data-stream
level may be well approximated by analytical formulae. At map and
\Cl\ level this is not always true, since the effect of the
scanning strategy combined with destriping and map-making
algorithms can not be modelled in a simple way. For these cases,
we have derived semi-analytical relations whose parameters are
calibrated through simulations.

The QRA process increases the RMS noise per pixel at the level
$\sim 1\%$\ for $\sigma/q \simeq 2$. The distribution of the RMS
induced by the QRA process depends on the scanning law and
resembles the distribution of white noise in the map. At low and
middle ecliptical latitudes, for {\sc Planck}/LFI the RMS
decreases toward the ecliptical poles scaling approximately as
$\sqrt{\sin \theta}$\ where $\theta$\ is the ecliptical
colatitude. The details of the distribution of the QRA additional
noise on the map depend on the frequency channel, the location of
the horn inside the focal plane and the scanning law. We have
derived the parameters that characterize this dependence for {\sc
Planck}/LFI. We find that the skewness and the kurtosis measured
all over the map are modified up to a few percent for $\sigma/q
\simeq 2$\, when the $1/f^\alpha$\ noise is subtracted with
destriping algorithms. Although not large, this effect is not
negligible compared to the sensitivity of the kurtosis index
recovery for a pure CMB sky. On the other hand, it is much less
relevant than residual foreground contamination which accurate
modelling turns to set the fundamental uncertainty on the recovery
CMB anisotropy high order moments.

Finally, we find that the quantization introduces a power excess,
$C_\ell^{ex}$, that, although related to the instrument and
mission parameters, is weakly dependent on the multipole $\ell$ at
middle and large $\ell$ and can be quite accurately subtracted.
For $\sigma/q \simeq 2$, the residual uncertainty, $\Delta
C_\ell^{ex}$, implied by this subtraction is of only
$\simeq$1--2$\%$ of the RMS uncertainty, $\Delta C_\ell^{noise}$,
on $C_\ell^{sky}$ reconstruction due to the noise power,
$C_\ell^{noise}$.

Only for $\ell \lsim 30$ the quantization
removal is less accurate; in fact, the $1/f$ noise features,
although efficiently removed, increase $C_\ell^{noise}$, $\Delta
C_\ell^{noise}$, $C_\ell^{ex}$ and then $\Delta C_\ell^{ex}$.
Anyway, at low multipoles $C_\ell^{\rm sky} \gg \Delta C_\ell^{\rm
noise} > \Delta C_\ell^{ex}$
and the uncertainty introduced by the
QRA effect is therefore in any case much less
than the unavoidable uncertainty
due the cosmic variance.

 \begin{acknowledgements}

We gratefully acknowledge K.M.~G\'orski and all the people
involved in the realization of the tools of HEALPix pixelisation.
We warmly thank ??? for having provided us with
the anisotropy map from thermal SZ effects
in clusters of galaxies and G. De Zotti and L. Toffolatti for
numberless discussions on foreground contamination.

 \end{acknowledgements}


 \appendix

 \section{calculation of $\alpha_{N_{\mathrm{s}}}$}\label{appendix:alphan}

 \TABLEAPPONE

The $\alphaN$\ coefficient in eq.~(\ref{eq:alpha:1sigma:noise}) is
defined as the ratio between the standard error for the QRA
process after \Ns\ averages $\tilde{\delta}_{N_{\mathrm{s}}}$ and
the RMS for the same process:

 \begin{equation}\label{eq:alphaN:def}
  \alphaN = \frac{\tilde{\delta}_{N_{\mathrm{s}}}}{\sigma_{N_{\mathrm{s}}}},
 \end{equation}

 \noindent
where \sigmaN\ is the RMS of the QRA error distribution after \Ns\
averages: $\uN(X)$, $\tilde{\delta}_{N_{\mathrm{s}}}$\ is obtained
from the definition of standard error:

  \begin{equation}\label{eq:tildeX:calc}
  \begin{array}{lll}
 I_{N_{\mathrm{s}}}(\deltatilde) & = & \int_{-\tilde{\delta}_{N_{\mathrm{s}}}}^{\tilde{\delta}_{N_{\mathrm{s}}}}
dt\, \uN(t) \\
   & = & \frac{1}{\sqrt{2\pi}} \int_{-1}^{+1} dt\; e^{-t^2/2} \approx 0.6827.\\
   \end{array}
 \end{equation}

 \noindent
We are interested in determine the central moments and the
integral of $\uN(\bar{\delta}_{N_{\mathrm{s}}})$\, i.e. to
determine the characteristic function of
$\uN(\bar{\delta}_{N_{\mathrm{s}}})$, from the characteristic
function for $S_{N_{\mathrm{s}}} = \Ns
\bar{\delta}_{N_{\mathrm{s}}}$. Where, $S_{N_{\mathrm{s}}}$\
represents the distribution of the {\em sum} instead of the {\em
average} of \Ns\ random variables. Taking in account that both
distributions are even and for $\Ns=1$\ they coincide with the
top-hat distribution: $u(x) = 1$, for $|x| \le 1/2$, or 0 for $|x|
> 1/2$, it is possible to demonstrate that the characteristic
function of $\uN(\bar{\delta}_{N_{\mathrm{s}}})$\ is:

 \begin{equation}\label{eq:characteristic:function}
 \Phi_{N_{\mathrm{s}}}(\omega)  = \left[
 \sync\left(\frac{\omega}{2N_{\mathrm{s}}}\right)
  \right]^{N_{\mathrm{s}}}.
 \end{equation}

 \noindent
where $\sync(x) = \sin(x) / x$. From
eq.~(\ref{eq:characteristic:function}) it is immediate to recover
the central moments of $\uN(\bar{\delta}_{N_{\mathrm{s}}})$. In
particular it is relevant to note that all the odd moments are
null, while: $m_2 = \sigma_{\Ns}^2/q^2 = 1/(12\Ns)$\ and
$m_4 = (5\Ns - 2)/(240 \Ns^3) = \mathrm{kurtosis}/q^4$.

The evaluation of \deltatilde\ requires to solve
eq.~(\ref{eq:tildeX:calc}). This is obtained by the following
formula:

 \begin{equation}\label{eq:rounding}
 \begin{array}{ll}
         I_{N_{\mathrm{s}}}(\deltatilde) =
        &  R(\omegamax)
        +\\
        &\\
        &\frac{4 \deltatilde}{\pi} \int_{0}^{\omega_{\mathrm{max}}}
        \,
          \sync\left(\omega \deltatilde \right)
                \left[
          \sync\left(\frac{\omega}{2 N_{\mathrm{s}}} \right)
                 \right]^{N_{\mathrm{s}}} \, d\omega \,
                 \, ,
 \end{array}
 \end{equation}

 \noindent
where $\omegamax$\ represents a cut-off in an otherwise
indefinitely extended integration range. The residual
$R(\omegamax)$\ is upper bounded by: $ R(\omegamax) \leq 4
(2N_{\mathrm{s}})^{N_{\mathrm{s}}}  / \left(\pi  \Ns
\omega_{\mathrm{max}}^{N_{\mathrm{s}}}\right)$\ which allows to
determine $\omegamax$\ once the absolute accuracy of integration
$\epsilon$\ is given:

 \begin{equation}\label{eq:omegamax:epsilon}
 \omegamax =
 2\Ns
 \left[
 \frac{
 4
 }{
  \pi \epsilon N_{\mathrm{s}}
 }
 \right]^{1/N_{\mathrm{s}}}.
 \end{equation}

 \noindent
Note that it would be also possible to solve
eq.~(\ref{eq:tildeX:calc}) through direct integration of
$\uN(\bar{\delta}_{N_{\mathrm{s}}})$, but this is not a
satisfactory method since the integral converges slowly for
$\Ns>10$.

 %

Values of \alphaN\ as a function of \Ns\ from
equations~(\ref{eq:alphaN:def})~and~(\ref{eq:tildeX:calc}) are
tabulated, with four digits of accuracy and for $\Ns = 1$, 2, 3,
4, 5, 6, 10, 15, 30, 60, in the last column of
Tab.~\ref{tab:app:one}. While the values of $\deltatilde$, and
\sigmaN\ are tabulated in the second and the third columns of the
table respectively. For $\Ns \ge 30$\ the following relation
holds:

 \begin{equation}\label{eq:alphan:fit}
   \log_{10}(\alphaN-1) \approx -0.923359 \cdot \log_{10}\Ns
   -1.129003.
 \end{equation}

 \noindent
with an accuracy compatible with the accuracy of the tabulated values.


 \section{the ADC quantization problem}\label{appendix:ADC:quantization}

In this paper we have extensively discussed the effect of the
software quantization of the data before lossless compression.
This quantization is only the last step for the on-board
data-processing which may affect the data quality, being the data
acquired by the on-board computer by an analog to digital
converter (ADC) which introduces at least three sources of noise:
$i)$ the intrinsic ADC quantization noise proportional to the ADC
quantization step \qADC; $ii)$ the read-out noise
$\sigma_{\mathrm{RON}}$; $iii)$ non linearities due to residual
differences in \qADC\ as a function of the input value seen by the
ADC. Detailed analyses of these problems are outside the scope of
this paper and will be addressed in forthcoming works. However, in
the context of this study on the signal digitisation effect it is
relevant to discuss the conditions under which the ADC
quantization effect is not critical compared to the software
quantization effect.

The impact of the ADC quantization depends on the steps of the
on-board data processing. Assuming $(a)$ to acquire separately the
sky and the reference signal with an ideal ADC (i.e. considering
negliglible the sources of noise $i)$ and $ii)$), $(b)$  to have a
perfectly balanced radiometer (i.e. both the sky and load
detectors have the same weight), $(c)$ the software quantization,
and then the total QRA error, to be independent from the ADC
quantization, and $(d)$ to have $\Nh$ ADC samples that are
averaged to obtain a given instrumental sample (i.e. the {\em
hardware} sampling rate is \Nh\ times higher than the {\em
software} sampling rate), then the global RMS quantization error,
$\sigma_g$, including both hardware and software quantization, is
given by:


 \begin{equation}\label{eq:two-steps:qerror}
  \sigma_g = \sigma_q
              \sqrt{
               \frac{2}{\Nh}
               \left( \frac{\qADC}{q} \right)^2
               + 1
              } \, ,
 \end{equation}

\noindent where $\sigma_q$ is the RMS of the software quantization
error. This relation is substantially correct at the level of TODs
as well as at the level of the maps, since the repeated
observation of the sky pixels and the averaging determined by the
scanning strategy operate in the same way in the case of data
affected by software quantitazion alone or by hardware plus
software quantization, analogously to the considerations of
Sect.~2.1. Clearly, $\sigma_g$ is larger than the expected QRA
error reported by eq.~(\ref{eq:qnoise:general}) which represents a
very good approximation in the limit $q \gg \qADC$ and/or for
large values of $\Nh$.

In order to understand the conditions under which the ADC
quantization effect is not critical compared to the software
quantization effect we focus on the impact on the angular power
spectrum estimation. At middle and large multipoles the power
spectrum of the global and the software quantization errors are
proportional to the square of the corresponding RMS quantization
error, i.e.:

 \begin{equation}
{ C^{\mathrm{\mathrm{QRA,g}}}_{l} - C^{\mathrm{\mathrm{QRA}}}_{l}
\over C^{\mathrm{\mathrm{QRA}}}_{l} } \simeq {
\sigma_g^2-\sigma_q^2 \over \sigma_q^2 } \, .
 \end{equation}

By following an approach analogous the that described in the
second part of Sect.~4, we require that

 \begin{equation}
{ C^{\mathrm{\mathrm{QRA,g}}}_{l} - C^{\mathrm{\mathrm{QRA}}}_{l}
\over C^{\mathrm{\mathrm{QRA}}}_{l} } \lsim  {1 \over \lambda} {
\mathrm{RMS}[C^{\mathrm{\mathrm{QRA}}}_{l}] \over
C^{\mathrm{\mathrm{QRA}}}_{l} } \, .
 \end{equation}

Differently from the case discussed in Sect.~4, we have now to
search for a law, $L_\ell$, which gives at each $\ell$ the minimum
value among the ratios
$\mathrm{RMS}[C^{\mathrm{\mathrm{exc}}}_{l,60}] /
C^{\mathrm{exc}}_{l,60}$ found for the set of simulations
considered here. We find that the approximate law

 \begin{equation}\label{eq:upplim:appendix}
L_\ell \simeq 0.46 \ell^{-0.5}
 \end{equation}

\noindent is a quite good approximation for the whole range of
multipoles relevant here.

By using the above relation between $\sigma_g$ and $\sigma_q$ and
this expression for $L_\ell$, the above condition is satisfied
provided that

\begin{equation}
{1 \over \Nh} \left( {\qADC \over q} \right)^2 \lsim {0.23
\ell^{0.5}  \over \lambda} \, .
\end{equation}

In the case of {\sc Planck}/LFI, $\Nh \sim 50$ and even in the
worst case of $q \simeq 0.5$~mK and $\qADC \simeq 1.2$~mK, this
condition is satisfied at $\ell \gsim 30$ with a value of $\lambda
\simeq 20$, which corresponds to a deviation from the value of
expectation of the angular power spectrum of pure software
quantization error of only few per cent of its RMS. Of course,
given an accurate description of the ADC quantization we can
easily include it in the data analysis. Anyway, this estimate
shows that its effect is typically very small.

Finally, it is worth to note that just replacing $2/\Nh$\ with
$1/\Nh$\ in eq.~(\ref{eq:two-steps:qerror}) accounts for a
different conceptual acquisition scheme where the sky and
reference signals are differenced before to perform the ADC
quantization. This scheme has a better propagation of the
quantization error than the scheme assumed in
eq.~(\ref{eq:two-steps:qerror}), implying that the above condition
can be satisfied with a value of $\lambda$ two times larger.

 \section{Statistical moments of simulated sky maps}\label{appendix:mom_maps}


We report here a comprehensive tabulation of the statistical
moments of the CMB simulated map adopted in this work compared
with the statistical moments of simulated maps of the most
relevant Galactic and extragalactic foregrounds. Our maps of
Galactic foregrounds are simulated according to Maino et al.
(2002), that of  extragalactic source fluctuations according to
the model by Toffolatti et al. (1998), while for thermal
Sunyaev-Zeldovich effects (SZ) from clusters of galaxies we have
used the template available at the MPI web site ???. It is infact
interesting to compare the modification on the statistical moments
of a pure CMB anisotropy sky due to the digisation effect with
that due to the foreground contamination.

The HEALPix scheme (G\`orski et al. 1998) is here adopted and different
resolutions, identified by the parameter $N_{\mathrm{side}}$ ($12\times N_{\mathrm{side}}^2$
is the number of pixels in the map),  are considered.

We consider here maps with and without beam convolution (FWHM of $33'$ as
appropriate to the LFI 30~GHz channels).
The input maps have been first simulated at $N_{\mathrm{side}} = 1024$
without beam convolution. Convolution and, possibly, degradation are then applied.

We report the statistical moments referring separately to each component
and to combination of CMB and foreground maps, by including or not masks
to exclude regions at low Galactic latitudes and/or high signal pixels,
as indicated in the tables. In fact, the
regions at low Galactic latitudes have to be avoided in CMB anisotropy analysis
as well as pixels significantly contaminated
by SZ effects or extragalactic sources have to be previously detected and removed.
More precisely, for uniformity reasons, we apply the same 5$\sigma$-clipping
threshold, $\sigma$ being the rms of CMB fluctuations at the considered
resolution for the corresponding case (including or not beam convolution),
independently of the considered kind of map composition.

Note that, while in the cases of unconvolved maps with large skewness and kurtosis
indices the convolution decreases both these estimators,
in the cases of unconvolved maps with moderate or small skewness and kurtosis
indices the convolution may produce a weak increase of these estimators
because the higher order moments are relatively less reduced than
the variance.


 \end{document}